\documentclass[journal]{IEEEtran}
\ifCLASSINFOpdf
\else
\fi
\usepackage{amsmath,amssymb,amsfonts}
\usepackage{algorithmic}
\usepackage{url}
\usepackage{acronym}
\acrodef{pmu}[PMU]{Phasor Measurement Unit}
\acrodef{wls}[WLS]{Weighted Least Squares}
\acrodef{wmr}[WMR]{Weighted Measurement Residual}
\acrodef{fdla}[FDLA]{Fault Detection and Localization Algorithm}
\acrodef{fdcla}[FDCLA]{Fault Detection and Cluster Localization Algorithm}
\acrodef{fdl}[FDL]{fault detection and localization}
\acrodef{ufc}[UFC]{Unlocalizable Fault Cluster}
\acrodef{der}[DER]{Distributed Energy Resources}
\acrodef{adn}[ADN]{Active Distribution Network}
\acrodef{dn}[DN]{Distribution Network}
\acrodef{mv}[MV]{Medium Voltage}
\acrodef{se}[SE]{State Estimator}
\acrodef{ba}[BA]{Basic Algorithm}
\acrodef{zsic}[ZSIC]{zero-sequence injected current}
\acrodef{feg}[fEG]{fictitious extended grid}
\acrodef{ufc2}[UFC*]{re-defined Unlocalizable Fault Cluster}
\acrodef{opa}[OPA]{Optimal Positioning Algorithm}
\acrodef{tso}[TSO]{Transmission System Operator}
\acrodef{tn}[TN]{Transmission Network}
\acrodef{fl}[FL]{Fault Localization}
\acrodef{dso}[DSO]{Distribution System Operator}

\usepackage{siunitx}
\DeclareSIUnit \voltampere { VA }
\DeclareSIUnit \wh { Wh }
\DeclareSIUnit \rad { rad }

\def \acc {\`}

\newtheorem{lemma}{Lemma}
\newtheorem{theorem}{Theorem}

\newtheorem{definition}{Definition}

\usepackage[ruled,vlined]{algorithm2e}
\usepackage{graphicx}
\usepackage{graphics}
\usepackage{multirow}
\usepackage{subcaption}

\usepackage{tikz}
\usetikzlibrary{shapes.geometric}
\usetikzlibrary{shapes.arrows}
\usepackage{array}

\usepackage{courier}

\IEEEoverridecommandlockouts
\newcommand\copyrighttext{%
  \footnotesize
  \centering\copyright~2022 IEEE. Personal use of this material is permitted. Permission from IEEE must be obtained for all other uses, in any current or future media, including reprinting/republishing this material for advertising or promotional purposes, creating new collective works, for resale or redistribution to servers or lists, or reuse of any copyrighted component of this work in other works. \\ Published on the IEEE Transactions on Power Systems, doi: 10.1109/TPWRS.2022.3165685.}
\newcommand\copyrightnotice{%
\begin{tikzpicture}[remember picture,overlay]
\node[anchor=south,yshift=0pt] at (current page.south) {\setlength{\fboxrule}{0pt}\fbox{\parbox{\dimexpr\textwidth-\fboxsep-\fboxrule\relax}{\copyrighttext}}};
\end{tikzpicture}%
}

\begin{document}
\title{Fault Detection and Localization in Active Distribution Networks using Optimally Placed Phasor Measurements Units}

\author{Francesco~Conte,~\IEEEmembership{Senior Member,~IEEE,}
        Fabio~D'Agostino,~\IEEEmembership{Member,~IEEE,}
        Bruno~Gabriele,~\IEEEmembership{Student Member,~IEEE,}
        Giacomo-Piero~Schiapparelli,~\IEEEmembership{Member,~IEEE,}
        and~Federico~Silvestro,~\IEEEmembership{Senior Member,~IEEE}% <-this % stops a space
\thanks{F. Conte, F. D'Agostino, B. Gabriele, G.-P. Schiapparelli, F. Silvestro are with the Department of Electrical, Electronics and Telecomunication Engineering and Naval Architecture, Universit\acc{a} degli Studi di Genova, Via all'Opera Pia, 11a, I-16145, Genova, Italy, e-mail: fr.conte@unige.it, fabio.dagostino@unige.it, bruno.gabriele@edu.unige.it, giacomo-piero.schiapparelli@edu.unige.it, federico.silvestro@unige.it }% <-this % stops a space

\thanks{F. Conte is also with Campus Bio-Medico University of Rome, Engineering Faculty, Via Alvaro del Portillo, 21, I-00128, Roma, Italy, e-mail: f.conte@unicampus.it}% <-this % stops a space
}
\IEEEaftertitletext{\copyrightnotice\vspace{1.1\baselineskip}}
\maketitle

\begin{abstract}
This paper introduces an algorithm able to detect and localize the occurrence of a fault in an Active Distribution Network, using the measurements collected by \acp{pmu}. First, a basic algorithm that works under the assumption that all grid buses are equipped with a \ac{pmu} is designed. Then, formal observability conditions that allow detection and localization with a reduced number of \acp{pmu} are provided. Based on these conditions, the algorithm is extended to perform correctly when not all network buses are monitored. Moreover, an Optimal Positioning Algorithm, always based on the observability conditions, is designed. This algorithm allows the user to customize the fault localization resolution. The approach is validated through simulations carried out on a benchmark active distribution network.  
\end{abstract}


\IEEEpeerreviewmaketitle

%%%%%%%%%%%%%%%%%%%%%%%%%%%%%%%%%%%%%%%%%%%%%%%%%%%%
\section{Introduction}
%%%%%%%%%%%%%%%%%%%%%%%%%%%%%%%%%%%%%%%%%%%%%%%%%%%
\IEEEPARstart{T}{he} huge deployment of \acp{der} has severely transformed the power system. Some of the traditional assumptions used to design grid management and control procedures are progressively loosing their validity. In particular, \acp{dn}, that have became ``active'', need to be operated in a very different way with respect to their original design~\cite{Bak:2015}. Indeed, the presence of \acp{der}, yields non-unidirectional power flows, which have a significant impact on the protection scheme behaviour. Moreover, classical equivalent models, that treat \acp{dn} as passive loads elements require to be updated in order to provide to the \ac{tso} with a suitable level of observability of the distribution system \cite{Yamashita:2012}. Observabilty can be enhanced also by installing distributed real-time monitoring devices such as \acp{pmu} and smart meters. 

Recent literature studies the possibility of merging protection devices, distributed measurements and communication systems, in order to asses in real-time the distribution system state and take decisions based on a complete view of the grid operating conditions. Such an approach opens to a plethora of applications both in operation and planning \emph{e.g.} state estimation \cite{Milano:2016,Conte:2020,Adinolfi:2013}, system and load modelling \cite{Zali:2013,Resende:2013,Conte:2019}, event detection and localization, and optimal positioning of measurement devices and switches \cite{Liu:2012}.

Within this context, the present paper focuses on the problem of \ac{fdl} in \acp{adn} using the measurements provided by \acp{pmu}. The goal is to detect the occurrence of a short-circuit in the \ac{dn} and state in which portion of the grid it has happened. 

\subsection{Literature overview}
In \cite{Pignati:2016} \ac{fdl} for \acp{adn} is realized using \ac{wls} \acp{se} and placing \acp{pmu} at each bus of grid, allowing the localization of the faulted line. In this work, the fault is also ``characterized'', \emph{i.e.} to state which phases have been faulted. In \cite{Farajollahi:2017}, \ac{fdl} is designed for radial passive \acp{dn}, allowing the localization of a fault between a couple of \acp{pmu}. This technique uses pseudo-measurements that introduce uncertainty in the provided results. The method proposed in \cite{Ardakanian:2017} allows \ac{fdl} in radial passive \acp{dn}; here, \acp{pmu} are placed at each bus of the grid. In \cite{Usman:2018} only localization is realized for radial \acp{adn} by placing \acp{pmu} at the end of each feeder and at all buses with \acp{der}. 

In \cite{Majidi:2017}, \ac{fdl} is carried out  for \acp{adn} by placing the \acp{pmu} within the grid buses, following an optimal positioning criteria. With this method, localization results are approximated. In \cite{Gholami:2019}, \ac{fdl} is realized for \acp{adn} by placing \acp{pmu} only to buses with \acp{der}. In this paper, pseudo-measurements are used and localization regards the identification of the faulted portion of the grid. In \cite{Jamei:2018} and \cite{Jamei:2020}, a method for \ac{fdl} in \acp{adn} is proposed, by placing \acp{pmu} at the point of connection with the main grid, at all buses with \acp{der}, and at a set of buses defined by optimal positioning. In this case, localization results are approximated, but estimation errors are characterized by a probabilistic model.

Some interesting approaches for \ac{fl} by \acp{pmu}' measurements are proposed also for \acp{tn}. This is the case of \cite{Lien:2006}, \cite{Golshan:2018} and \cite{Alexopoulos:2016}. In these works, the fault location is intended as the position of a fault along a transmission line.

A problem correlated to \ac{fdl} is the optimal positioning of \acp{pmu}. Indeed, the number of buses in a \ac{dn} can be very high and to install a \ac{pmu} at each bus may result not achievable from the economical point of view. Therefore, many papers study methods to find the minimum number and the suitable positions of \acp{pmu}, in order to keep the level of observability required to carry out state estimation and/or \ac{fdl}. 

In \cite{Singh:2009, Singh:2011} and \cite{Pegoraro:2013}, three different optimization approaches are proposed to optimally place meters that measure bus voltages and lines power flows, with the objective of minimizing the state estimation error, exploiting also pseudo-measurements. In \cite{Shafiu:2005}, a heuristic method is introduced to place voltage meters to reduce the voltage estimation error in the non-monitored buses. In \cite{Damavandi:2015}, \ac{pmu} that measure only voltages are optimally placed to obtain a given level of state estimation accuracy, considering also different kind of network configurations. The paper \cite{Liu:2012} establishes an optimal trade-off between the deployment of smart meters and \acp{pmu} with the objective of obtain a given level of estimation accuracy and minimize the costs. The same approach is extended in \cite{Liu:2014} to guarantee robustness in $N-1$ conditions. The works \cite{Xygkis:2017} and \cite{Xygkis:2018} propose two methods for \acp{adn} to optimally place \acp{pmu} and smart meters with the objective of minimizing the state estimation error. These two approaches are based on formal probabilistic principles and also allow the use of pseudo-measurements. In \cite{Jamei:2018} and \cite{Jamei:2020} an optimization algorithm is proposed to place \acp{pmu} in order to carry out \ac{fdl} in \acp{adn}.

All the mentioned papers adopt probabilistic or heuristic optimization methods since the general objective is to minimize the error in a state estimation process, except for the two last references \cite{Jamei:2018,Jamei:2020}. Deterministic approaches, based on formal observability conditions can be mainly found for \acp{tn}, as happens, for example, in \cite{Abur:2004},  \cite{Gou:2008}, {\cite{Kavasseri:2011} and \cite{Pokharel:2009}}, where \acp{pmu} are placed by linear mixed-integer programming. In \cite{Lien:2006}, \cite{Golshan:2018} and \cite{Alexopoulos:2016}, \acp{pmu} are optimally placed in \acp{tn} to obtain faults observability, roughly defined as the possibility to state the position of a fault along a transmission line. Faults observability in \acp{tn} is assured also in \cite{Korkali:2013} and \cite{Liao:2009}, where optimal placement of voltage meters is carried out.

Recent literature proposes optimal placement algorithms based on formal observability conditions for \acp{dn}. In \cite{Biswas:2020} and \cite{Su:2019} \acp{pmu} are placed in a \ac{dn} by linear mixed-integer optimization, to obtain the grid observability and perform state estimation. In this paper, constrains that take into account the presence of zero injection buses and network configuration changes are included in the optimization. In \cite{Teimourzadeh:2019}, \acp{pmu} and smart meters are optimally placed in \acp{adn} to maintain observability at steady-state and contingencies conditions. The papers \cite{Dua:2019} and \cite{Dua:2021} propose an optimal \acp{pmu} placement method in reconfigurable \acp{dn} by multi-objective mixed-integer optimization. 

\subsection{Paper contributions}
The present paper introduces an algorithm for \ac{fdl} in radial \acp{adn}. The approach is inspired by the one of \cite{Pignati:2016}, where a series of \ac{wls} \acp{se} are carried out considering all possible network topologies, for each fault location (each line). Differently from \cite{Pignati:2016}, the objective is to realize \ac{fdl} without positioning \acp{pmu} at each bus, but looking for the minimum necessary number of such devices. The contributions of the paper are listed in the following. 
\begin{itemize}
    \item A \ac{fdla} able to detect, characterize and identify the line (localize) where a fault has occurred is introduced, assuming that all buses are monitored by a \ac{pmu}, similarly to \cite{Pignati:2016}. With respect to this paper, detection robustness is improved by using a double check and adopting different indices, one involving the \ac{wmr} and one involving the estimated zero-sequence injected currents. Moreover, localization is realized with a different criteria.
    \item Formal observability conditions are provided, that allow the \ac{fdla} to work correctly, by removing the assumption that all buses are monitored by a \ac{pmu}. Moreover, it is proved that, in this case, the grid results to be partitioned into connected lines clusters, within which it is not possible to distinguish in which line the fault has occurred. Theoretical results are  specifically referred to the objective of detecting the fault and to localize the faulted portion of a radial \acp{adn}. This marks a difference with the existing literature, which provide fault observability results mainly for meshed \acp{tn} (\cite{Lien:2006,Golshan:2018,Alexopoulos:2016,Kavasseri:2011,Pokharel:2009,Korkali:2013,Liao:2009}), and with the objective of stating the exact position of the fault along the transmission line.    
    \item Based on these theoretical results, the \ac{fdla} is transformed into a \ac{fdcla}, that, given a position of the \acp{pmu} in the grid, recognizes the lines cluster where the fault has occurred.
    \item A \acp{pmu} \ac{opa} is developed. It is based on the observability conditions previously provided. The advantage of the method is that the \ac{opa} can be customized to: obtain a desired localization resolution, in terms of number and composition of lines clusters; allow \ac{fdl} with any possible network topology, in the case of reconfigurable \acp{adn}.
\end{itemize}

The \acp{pmu} \ac{opa} and the \ac{fdcla} are validated by simulations carried out on a benchmark \ac{adn}. Preliminary results of this paper have been published in \cite{Conte:2021}.

The paper is organized as follows: Section~\ref{sec:ProblemStatement} reports the problem formulation; Section~\ref{sec:fdla} introduces the \ac{fdla}; Section~\ref{sec:observability} defines the theoretical observability and fault localizability conditions; Section~\ref{sec:fdcla} provides the \ac{fdcla}; Section~\ref{sec:OptimalPositioning} introduce the \ac{opa}; Section~\ref{sec:results} presents simulation results; Section~\ref{sec:conclusions} provides the paper conclusions.

\vspace{10pt}
\textit{Notation.} $\dot{\mathbf{U}}^{abc}$ is a triplet vector $\begin{bmatrix} \dot{U}^{a} & \dot{U}^{b} & \dot{U}^{c} \end{bmatrix}$. Real and imaginary components of a generic phasorial vector $\dot{U}$ are indicated as $U_{re}$ and $U_{im}$. $b_i$ indicates the $i$-th bus of the grid; $\rho(b_i)$ is the degree of $b_i$, defined as the number of buses connected with $b_i$. $|\mathcal{S}|$ is the cardinality of a given set $\mathcal{S}$. 

%%%%%%%%%%%%%%%%%%%%%%%%%%%%%%%%%%%%%%%%%%%%%%%%%%%%%%%%
\section{Problem Statement} \label{sec:ProblemStatement}
%%%%%%%%%%%%%%%%%%%%%%%%%%%%%%%%%%%%%%%%%%%%%%%%%%%%%%%%
The objective of this work is to detect and localize a fault event occurring in a \ac{mv} \ac{adn}. The following hypotheses are supposed to hold true: 
\begin{itemize}
    \item[${h}_1$:] the \ac{adn} is radial and composed by $n$ buses and $m$ lines;
    \item[${h}_2$:] the network admittance matrix $Y$ is known;
    \item[${h}_3$:] $d\leq n$ buses are monitored via \acp{pmu}, and $n-d$ buses are not; $\mathcal{M}_b$ is the set of the indices of the $d$ monitored buses; $\mathcal{N}_b$ is the set of the indices of the $n-d$ non-monitored buses.
    \item[${h}_4$:] measurements acquired at each monitored bus $b_i$ ($i\in\mathcal{M}_b$) are: three phase-to-ground voltages $\dot{\mathbf{V}}_i^{abc}$ and three phase injected currents $\dot{\mathbf{I}}_i^{abc}$; measurements are collected with given sampling time $\Delta t$.
\end{itemize}

Based on these hypotheses, state vector $x$ is defined as
\begin{align}
& x=\begin{bmatrix}   \mathbf{V}^{abc}_{1,re} & \cdots & \mathbf{V}^{abc}_{n,re} &  \mathbf{V}^{abc}_{1,im}& \cdots & \mathbf{V}^{abc}_{n,im} \end{bmatrix}^\top \in \mathbb{R}^N, 
\end{align}
with $N = 6n$, and measurements vector $z$ is defined as
\begin{equation}\label{eq:measurementVector}
z=[z_V,z_I]^\top \in \mathbb{R}^{D}, \quad D=12d,
\end{equation}
with
\begin{align}
	z_V & = 
	\begin{bmatrix} \cdots & \mathbf{V}^{abc}_{re,i} & \cdots &\mathbf{V}^{abc}_{im,i} & \cdots 
    \end{bmatrix},  \quad i \in \mathcal{M}_b, \label{eq:zV} \\
	z_I&=
	\begin{bmatrix}
	\cdots & 
    \mathbf{I}^{abc}_{re,i} & \cdots  &\mathbf{I}^{abc}_{im,i} & \cdots 
    \end{bmatrix}, \quad i \in \mathcal{M}_b. \label{eq:zI}
\end{align}

The relation between $x$ and $z$ is given by
\begin{equation}\label{eq:output equation}
    z = Hx + v
\end{equation}
where $H$ is a $D\times N$ matrix and $v$ is the measurement noise, introduced by \acp{pmu}. $v$ is supposed to be zero-mean and characterized by a known covariance matrix $R$ \cite{Milano:2016}.

Matrix $H$ has the following form:
\begin{equation}
H = \begin{bmatrix}
    H_V \\
    H_I
\end{bmatrix}
\end{equation}
where $H_V\in\mathbb{R}^{6d \times N}$ relates the state vector $x$ with the component $z_V$ of $z$, and $H_I\in\mathbb{R}^{6d \times N}$ relates $x$ with the component $z_I$ of $z$. Thus, $H_V$ is composed by zeros and ones that select the voltages directly measured by \acp{pmu}, whereas $H_I$ is computed using the network admittance matrix $Y$, by removing from the following matrix
\begin{equation}
\mathcal{H}_I = \begin{bmatrix}
    Re\{Y\} & -Im\{Y\} \\
    Im\{Y\} & Re\{Y\} 
\end{bmatrix},
\end{equation}
the rows corresponding to non-monitored buses.

The objectives of the paper are:
\begin{enumerate}
    \item to develop an algorithm that, using the available \acp{pmu} measurements, is able to: \emph{1.a)} \textit{detect} the occurrence of a fault; \emph{1.b)} \textit{localize} the fault; \emph{1.c)} \textit{characterize} the fault, \textit{i.e.} state in which phases it has occurred; 
    \item to define the minimum number $d$ and the optimal positions of \acp{pmu} along the grid, that allow the algorithm to work correctly and obtain a desired fault localization resolution.
\end{enumerate}

%%%%%%%%%%%%%%%%%%%%%%%%%%%%%%%%%%%%%%%%%%%%%%%%%%%%%%%%
\section{Fault Detection and Localization Algorithm}\label{sec:fdla}
%%%%%%%%%%%%%%%%%%%%%%%%%%%%%%%%%%%%%%%%%%%%%%%%%%%%%%%%
In this section, we introduce an algorithm able to detect and localize a fault in the hypothesis that each bus is monitored, \textit{i.e.} $d=n$. The approach adopted in this paper is similar to the one of \cite{Pignati:2016}. The idea is that a fault on a line causes a sudden addition of one \textit{virtual bus}, placed between two real buses, that absorbs the fault current. Therefore, $m+1$ parallel \ac{wls} \acp{se}  \cite{Milano:2016} are realized, each returning, at any measurement (discrete) time step $t=0,1,2,\ldots$, the estimate $\hat{x}^k_t$, $k=0,1,\ldots,m$. Estimate $\hat{x}^{0}_t$ is obtained by applying \ac{wls} without adding any virtual bus, as follows: 
\begin{equation}\label{eq:wls0}
\hat{x}^{0}_t=\left({H}^\top R^{-1} H\right)^{-1} {H}^\top R^{-1} z_t.
\end{equation}

Estimates $\hat{x}^{k}_t$, with $k>0$, are computed by applying the \ac{wls} equation to the network extended with a virtual bus placed in the middle of line $k$. In particular, for all $k=1,2,\ldots,m$: 

\noindent \emph{a)} state vector $x$ is extended with the voltage of the virtual bus $b_{n+1}$:
\begin{equation}
\label{eq:extended_state_vector}
x^k=\begin{bmatrix}   \mathbf{V}^{abc}_{1,re} & \cdots & \mathbf{V}^{abc}_{n+1,re} & \mathbf{V}^{abc}_{1,im}& \cdots &  \mathbf{V}^{abc}_{n+1,im}\end{bmatrix}^\top; 
\end{equation}
\emph{b)} the admittance matrix of the extended network $Y^k$ is computed and used to obtain the corresponding measurement matrix $H^k$; 

\noindent \emph{c)} estimate $\hat{x}^k_t$ is computed as follows:
\begin{equation}\label{eq:wlsk}
\hat{x}^{k}_t=\left({H^k}^\top R^{-1} H^k\right)^{-1} {H^k}^\top R^{-1} z_t.
\end{equation}

To evaluate the estimate accuracy, each \ac{se} is associated with the \ac{wmr}, defined as follows:
\begin{equation}\label{eq:wmr}
w^k_t=\sqrt{(z_t-H^k\hat{x}^k_t)^\top R^{-1}(z_t-H^k\hat{x}^k_t)}.
\end{equation}

\subsection{Detection}
Similarly to \cite{Pignati:2016}, the idea is that in normal operating conditions, with no fault occurrence, this procedure will return $m+1$ estimates with comparable \acp{wmr}, because all \acp{se} are using a correct model of the grid topology, in which the virtual bus is coherently estimated to do not absorb any current. Differently, if a fault happens on line $\alpha$, only the $\alpha$-th \ac{se} will be computed according to a grid topology close to real one. This implies that all \acp{wmr}, excepting for the $\alpha$-th, will sharply increase. In particular, the $0$-th \ac{wmr} will always increase, since the grid topology model used by the corresponding \ac{se} does not include any virtual bus.

According to this idea, a fault can be detected by checking if an anomalous variation of $w^0$ occurs, \textit{i.e.} by verifying at each measurement time step $t$ if 
\begin{equation}\label{eq:detection_rule_wmr}
    |(w^0_t-w^0_{t-1})-\mu_{\Delta w^0}|>th_w, 
\end{equation}
where $th_w$ is the detection threshold and $\mu_{\Delta w^0}$ is the mean value of the sequence $\{w^0_t-w^0_{t-1}\}_t$. These quantities can be set by estimating numerically the 
distribution of the variations $w^0_t-w^0_{t-1}$, using a consistent set of samples, collected during no-fault operating conditions. The detection threshold $th_w$ can be set equal to the 99.9$\%$ percentile of such a distribution. In this way, to satisfy \eqref{eq:detection_rule_wmr} will be a strongly unlikely event.   

As mentioned before, the use of \eqref{eq:detection_rule_wmr} to detect a short-circuit is based on the fact that in a faulted scenario the model on which the 0-th \ac{se} is based is wrong. However, when the short-circuit current is low, as it may happen in the 1-phase fault case, especially when neutral is connected to ground by Peterson coil, the model is only slightly wrong and the variation of $w^0$ risk to be comparable to the ones that usually arise because of noise. This means that the use of condition \eqref{eq:detection_rule_wmr} can cause a false negative in detecting a 1-phase fault. 

Therefore, a second fault detection check is adopted in this paper. It is well known that a single-phase to ground short-circuit implies a value of the zero-sequence injected current different from zero. This current should be ideally present in the fault location and in the bus where the neutral is connected to the ground. Given the voltages estimates $x_t^k$, $k=1,2,\ldots,m$ of the \acp{se} that assume the presence of virtual buses, we can compute the estimates of injected currents, using the corresponding output matrix, \textit{i.e.} $\hat{z}^k_{I,t}=H_I^k \hat{x}^k_t$. From $\hat{z}^k_{I,t}$, we can extract the current injected at the bus where the neutral is connected to the ground and compute the amplitude of the relevant zero-sequence injected current, indicated in the following as $\hat{I}_{0ng,t}^k$. Therefore, the following condition can be used to detect a single-phase short circuit:
\begin{equation}\label{eq:detection_rule_zero_sequence}
    \max_k\{\hat{I}_{0ng,t}^k\}>th_{0ng}. 
\end{equation}
As for condition \eqref{eq:detection_rule_wmr}, the detection threshold $th_{0ng}$ can be set equal to the 99.9$\%$ percentile of the distribution of the quantities $\max_k\{\hat{I}_{0ng,t}^k\}$, that can be estimated numerically by using a consistent set of samples (\textit{i.e.} $>$ 1000) collected during no-fault operating conditions.

To conclude, the fault detection is operated by verifying first \eqref{eq:detection_rule_wmr} and then, if it is not satisfied, by checking \eqref{eq:detection_rule_zero_sequence}. If one of these two conditions hold true, a fault is detected and the current time is marked as the \textit{fault time} $t_F$. 

\subsection{Localization}\label{ssec:localization}
Once a fault has been detected, localization is carried out using all \acp{wmr} $w^k$ of the \acp{se} that assume the presence of virtual buses. For each $k=1,2,\ldots,m$, the average value $\mu_{w}^k$ of the sequence $\{w^k_t\}_t$, in no-fault conditions, is supposed to be available. Such a quantity can be easily computed numerically, by using a consistent set of samples (\textit{i.e.}$>$1000) collected during normal operating conditions. Therefore, the faulted line $k_F$ is identified by selecting the \ac{wmr} with the minimum variation from its no-fault average value $\mu_{w}^k$ at the faulted time $t_F$, \textit{i.e.}:
\begin{equation}\label{eq:localization_rule}
    k_F = \arg \min_k \{ |w_{t_F}^k- \mu_{w}^k| \}.
\end{equation}

Notice that, differently, in \cite{Pignati:2016} localization is carried out by selecting the minimum \ac{wmr} at the fault time instead of the minimum variation. This coincides \eqref{eq:localization_rule} if $\mu_{w}^k$ are equal each others. Actually, we verified that this is approximately true only when all buses are monitored ($d=n$), whereas, when $d<n$ the $m$ \acp{se} returns \acp{wmr} with different average values. Thus, especially when short-circuits currents are low, selecting the minimum \ac{wmr} can compromise localization. Indeed, if noise is high and short-circuit current is low, at the fault time it can happen that one of the wrong \acp{se} significantly increases from its average value, but not enough to overcome the \ac{wmr} of the right \ac{se} which stays around its own average value.

This motivates the use of \eqref{eq:localization_rule}, which allows the fault localization immediately after the detection (at the same measurement time step). Obviously, because of noise, localization failures cannot be fully avoided. A way to be more robust with respect to noise is to use more than one measurement after the fault, accepting a delay $\delta$ in the localization result. In this case, localization is carried out at time step $t_F+\delta$ by considering the average value of the \acp{wmr} after the fault within a time window of length $\delta+1$, \textit{i.e.}:
\begin{equation}\label{eq:localization_rule_delay}
    k_F = \arg \min_k \left\lbrace \left|\frac{1}{\delta+1}\sum_{\tau=0}^{\delta}w_{t_F+\tau}^k- \mu_{w}^k\right| \right\rbrace.
\end{equation}

Robustness with respect noise will be higher as higher is the delay $\delta$. The choice of the value of $\delta$ depends on why the fault location is required. For example, if it is required to activate any protection, localization should be fast and, depending on the case, a maximum delay of 1 or 2 sampling times (20-40 ms) can be acceptable.
Relation \eqref{eq:localization_rule_delay} can be also used as a \textit{confirmation} to be processed after having executed \eqref{eq:localization_rule}. In general, we consider $\delta$ as a settable parameter of the \ac{fdla}. Notice that if $\delta=0$ \eqref{eq:localization_rule_delay} coincides with \eqref{eq:localization_rule}.

\begin{algorithm}[t] 
 \caption{\ac{fdla}}
\SetAlgoLined
\DontPrintSemicolon
\KwData{$\delta,\mu_{\Delta w^0},th_w,th_{0ng},\gamma,th_v,R,H^k,\mu_{w}^k$}
\textit{FaultDetected} $\leftarrow$ 0;\;
\textbf{for all} measurement time step $t$ \textbf{do}\;
collect \acp{pmu} measurements $z_t$;\; 
compute $\hat{x}^{k}_t$ and $w^{k}_t$, $\forall k=0,1,\dots,m$;\;
[\textit{FaultDetected}, $t_F$] $\leftarrow$ \texttt{detectFault}($w^0_t$, $w^0_{t-1}$, $\hat{x}_t^k$);\;
\textbf{return} \textit{FaultDetected}, $t_F$;\;
\If{\textit{FaultDetected} \textbf{and} $t=t_F+\delta$}
{$k_F$ $\leftarrow$ \texttt{localizeFault}($\{w^0_{t_F+\tau}\}_{\tau=0,1,\ldots,\delta}$);\;
\textit{FaultedPhases} $\leftarrow$ \texttt{characterizeFault}($\hat{x}_{t_F}^{k_F}$);\;
\textit{FaultDetected} $\leftarrow$ 0;\;
\textbf{return} \textit{FaultedPhases, $k_F$};\;}  
\textbf{endfor}\;
\;
% Set Function Names
  \SetKwFunction{FD}{detectFault}
  \SetKwFunction{FL}{localizeFault}
  \SetKwFunction{FC}{characterizeFault}
% Detection
  \SetKwProg{Fn}{Function}{:}{}
  \Fn{\FD{$w^0_t$, $w^0_{t-1}$, $\hat{x}_t^k$}}{
\eIf{$|(w^0_t-w^0_{t-1})-\mu_{\Delta w^0}|> th_w$}{
   \textit{FaultDetected} $\leftarrow$ 1;\;
   $t_F \leftarrow t$;\;
   }{
   extract $\hat{I}_{0ng,t}^k$ from $\hat{z}^k_{I,t}=H_I^k \hat{x}^k_t,  \forall k=1:m$;\;
   \If{$\max_k\{\hat{I}_{0ng,t}^k\}> th_{0ng}$}
   {\textit{FaultDetected} $\leftarrow$ 1;\;
   $t_F \leftarrow t$;\;}
  }
  \KwRet \textit{FaultDetected}, $t_F$;\;
  }
  \;
% Localization
    \SetKwProg{Fn}{Function}{:}{}
  \Fn{\FL{$\{w^0_{t_F+\tau}\}_{\tau=0,1,\ldots,\delta}$}}{
$k_F \leftarrow \arg \min_k \left\lbrace \left|\frac{1}{\delta+1}\sum_{\tau=0}^{\delta}w_{t_F+\tau}^k- \mu_{w}^k\right| \right\rbrace$;\;
\KwRet $k_F$;\;
  }
  \;
  % Charaterization
    \SetKwProg{Fn}{Function}{:}{}
  \Fn{\FC{$\hat{x}_{t_F}^{k_F}$}}{
  extract $V_{vb}^F$ from $\hat{x}_{t_F}^{k_F}$;\;
extract $\hat{I}_{vb}^F$ from $H^{k_F} \hat{x}^{k_F}_{t_F}$;\;
$\hat{I}_{vb,max}^F\leftarrow \max \{\hat{I}_{vb}^F\}$;\;
\textit{FaultedPhases} $\leftarrow$ components in $\hat{I}_{vb}^F$ $>\gamma \cdot \hat{I}_{vb,max}^F$;\;
\If{(\# components in $\hat{V}_{vb}^F < th_{v})= 1$}
   {\textit{FaultedPhases} $\leftarrow$ component in $\hat{V}_{vb}^F < th_{v}$;\;}
\KwRet \textit{FaultedPhases};\;
  }
  \;
  \vspace{-12pt} 
\end{algorithm}

\subsection{Characterization}\label{ssec:characterization}
Once a fault has been detected and localized on line $k_F$, we can establish which phases have been faulted by computing from the selected best estimate $\hat{x}^{k_F}_{t_F}$ the amplitude of the voltages ${V}_{vb}^F$ and the currents ${I}_{vb}^F$ injected into the virtual bus. The estimate $\hat{V}_{vb}^F$ can be extracted directly from $\hat{x}^{k_F}_{t_F}$, whereas $\hat{I}_{vb}^F$ can be extracted from vector $H^{k_F} \hat{x}^{k_F}_{t_F}$. 

A double-check strategy is adopted. First, faulted phases are identified by non-zero injected currents at the virtual bus. Because of noise, all components in the estimated vector $\hat{I}_{vb}^F$ are never exactly equal to zero. Therefore, injected currents are labelled as non-zero if they overcome a threshold defined as $th_I = \gamma \hat{I}_{vb,max}^F$, where $\hat{I}_{vb,max}^F$ is the maximal component of $\hat{I}_{vb}^F$ and $\gamma$ is a coefficient close to zero (\textit{e.g.} $\gamma=0.2$). 

For 3- and 2-phases faults, this first check is effective in robustly characterizing the fault since fault currents are always significantly higher than zero. This is true also for one-phase faults with earthed neutral. Differently, when neutral is
compensated, fault currents may be very low and they risk being confused with the estimation error noise. 
This problem can be solved using $\hat{V}_{vb}^F$. Indeed, if the fault is 1-phase there will be a unique voltage at the virtual bus close to zero. Therefore, we first verify if there is a unique component in $\hat{V}_{vb}^F$ lower than a given threshold $th_{v}$ (\textit{e.g.} $th_{v}=0.05$ p.u.) and, if this holds true, the result obtained in the first check is overwritten. 

\vspace{10pt}
To summarize, the complete  pseudo-code of the \ac{fdla} is reported in Algorithm~1, where the three main procedures of fault detection, localization and characterization are implemented by the three different functions \texttt{\small detectFault}, \texttt{\small localizeFault}, and \texttt{\small characterizeFault}.

It is worth remarking that the proposed algorithm is intended to be integrated and not to substitute the existing protection schemes. The goal is to introduce higher-level functionalities and to overcome the limits of the state-of-the-art configurations. Indeed, while a large share of fault occurrences can be detected (protection tripping), localization is challenging and therefore selectivity is not always ensured, especially with the presence of distributed generation. In the proposed methodology, fault detection acts as a trigger for the fault localization task using distributed information. It allows \acp{dso} to identify the exact portion of the network affected by the fault, so that both fault handling and restoration process can be effective.

%%%%%%%%%%%%%%%%%%%%%%%%%%%%%%%%%%%%%%%%%%%%%%%%%%%%%%%%%%%%%%%%%
\section{Grid observability and fault localizability}\label{sec:observability}
%%%%%%%%%%%%%%%%%%%%%%%%%%%%%%%%%%%%%%%%%%%%%%%%%%%%%%%%%%%%%%%% 
The \ac{fdla} introduced in Section~\ref{sec:fdla}, does not generally works in the hypothesis that not all buses are monitored by \acp{pmu} (\textit{i.e.} $d<n$). First, \ac{wls} estimates \eqref{eq:wls0} and \eqref{eq:wlsk} cannot be implemented independently of the number and the position of \acp{pmu}, since this is possible only if the network \textit{observabilty} is kept \cite{Abur:2004}. Second, even if \eqref{eq:wls0} and \eqref{eq:wlsk} are computable, localization cannot correctly carried out using \eqref{eq:localization_rule} or \eqref{eq:localization_rule_delay} independently of the number and the position of \acp{pmu}. The latter is clarified in this section, where the conditions to keep the network observability with a reduced number of \acp{pmu} are also provided.

Following the definitions in \cite{Abur:2004}, an electrical network is observable if, in the no noise ideal case, it is possible to compute the voltages at all buses using the available measurements. In the formulation adopted in this paper, we have observability if and only if matrix $H$ is full row-rank. It is well known that, if this does not hold true, matrix $H^\top R^{-1}H$ in \eqref{eq:wls0} is not invertible and, thus, \ac{wls} cannot be used.      

If all buses are equipped with a \ac{pmu}, all nodal voltages are directly known ($H_V$ is the identity matrix in $\mathbb{R}^{6n}$), and the network is observable. However, our objective is to reduce as much as possible the number of \acp{pmu}. 

By applying simple electrotechnical computations it is easy to prove the following lemma.
\begin{lemma}\label{lem:1}
A radial grid with $n$ buses and $m$ lines is observable if, for all $i=1,2,\ldots,n$:
\begin{itemize}
\item[a.] if $\rho(b_i)>1$, $b_i$ has at most one adjacent non-monitored bus;
\item[b.] if $\rho(b_i)=1$ and $b_i$ is non-monitored, then $b_i$ is adjacent to a monitored bus.
\end{itemize}
\end{lemma}

Lemma \ref{lem:1} provides a sufficient condition that allows \eqref{eq:wls0} to be implemented. However, we need to implement also estimates \eqref{eq:wlsk}. Therefore, the observability of the $m$ networks extended with a virtual bus placed in the middle of each line is required. Indeed, this implies that matrices $H^k$ are full row-rank and, thus, ${H^k}^\top R^{-1} {H^k}$ are invertible. The following theorem provides the conditions to satisfy this requirement. 

\begin{theorem}\label{th:1}
Given a radial grid with $n$ buses and $m$ lines, the extended grid with $n+1$ buses and $m+1$ lines, obtained by adding the virtual bus $b_{n+1}$ in the middle of one of the $m$ lines is observable if and only if, for all $i=1,2,\ldots,n$:
\begin{itemize}
\item[a.] whatever given a couple of adjacent buses, at least one of the two is monitored;
\item[b.] if $\rho(b_i)=1$, $b_i$ is monitored.
\end{itemize}
\end{theorem}

Theorem \ref{th:1} has been provided and proved in \cite{Conte:2021}, thus the proof is not reported in the present paper for space lacking. 
      
In summary, Theorem~\ref{th:1} states that to obtain network observability, terminal buses must be monitored and there cannot be two adjacent non-monitored buses. 

With Theorem~\ref{th:1}, we have the conditions to make estimates \eqref{eq:wls0} and \eqref{eq:wlsk} implementable and correctly working.
Next Theorem~\ref{th:2} provides a key property of such estimates that makes a fault unlocalizable within specific portions of the grid by using \ac{wls} estimates. Such grid portions are defined as follows.

\begin{definition}\label{def:1}
An \ac{ufc} is a connected portion of the grid, defined as a subset of lines $\mathcal{C}\in\{1,2,\ldots,m\}$, such that one of the two following conditions a. or b. are satisfied:
\begin{itemize}
\item[a.] it is a unique line connecting two monitored buses or connecting a monitored bus with a non-monitored one;
\item[b.] it is a set of lines such that: 1) whatever given one line in $\mathcal{C}$, it connects a monitored bus to a non-monitored one; 2) it is connected to the rest of the grid only through non-monitored buses; 3) no more than one line in $\mathcal{C}$ is incident to non-monitored bus of degree $>$ 2.
\end{itemize}
\end{definition}

\begin{figure}[t]
  \centering
    \includegraphics[width=1\linewidth]{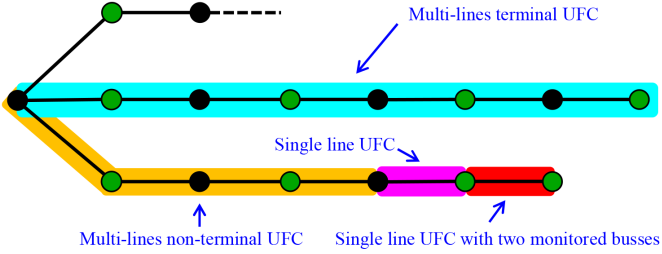}
  \caption{Types of \acp{ufc}. Green buses are monitored; black buses are not monitored.}
  \label{fig:clusters}
\end{figure}

As showed in Fig.~\ref{fig:clusters}, according to Definition~\ref{def:1}, four possible types of \ac{ufc} $\mathcal{C}$ can be found in a given grid. Two types are composed of a single line, one connecting two monitored buses, one connecting a monitored with a non-monitored bus. The two other types of \ac{ufc} are multi-lines and they can be \textit{non-terminal}, in the case that no line in $\mathcal{C}$ is incident at a terminal bus, or \textit{terminal}, when one line in $\mathcal{C}$ is incident at a terminal bus. 

\begin{theorem}\label{th:2}
Consider a radial grid with $n$ buses and $m$ lines and suppose that conditions a. and b. of Theorem~\ref{th:1} are satisfied. Whatever given an \ac{ufc} $\mathcal{C}$ and two estimates $\hat{x}^{i}$ and $\hat{x}^{\ell}$, computed by \eqref{eq:wlsk}, associated to virtual buses placed on two lines belonging to $\mathcal{C}$, then, for all $z\in\mathbb{R}^{D}$, $w^{i}=w^{\ell}$.
\end{theorem}

{The proof of Theorem~\ref{th:2} is reported in the Appendix section.} Theorem~\ref{th:2} implies that, using the \ac{fdla} introduced in Section~\ref{sec:fdla}, we will obtain a set of grid clusters associated to \ac{wls} estimates with identical \acp{wmr}. Therefore, an idea could be to compute a unique estimate $x^k$ for each \ac{ufc} and discover in which one the fault has occurred applying \eqref{eq:localization_rule} or \eqref{eq:localization_rule_delay}. Unfortunately, differently on what happens when all buses are monitored \cite{Pignati:2016}, in the present case, putting a virtual bus in the middle of one line within the correct \ac{ufc} does not necessarily result in the best representation of the faulted grid, excepting for the unlikely case of occurrence of the fault at the exact half of the line. 

In particular, this happens when the fault occurs within a single-line \ac{ufc}, with two monitored buses (the red one in Fig.~\ref{fig:clusters}). We verified that, in these cases, if the fault occurs sufficiently far from the middle of the line, the \ac{wmr} of any adjacent line could vary, at the fault time, less than the one associated to the right one, yielding rules \eqref{eq:localization_rule}-\eqref{eq:localization_rule_delay} to choose the wrong line. 
This can be explained by the fact that both the \acp{se} associated to the correct and the wrong adjacent \acp{ufc} use an approximated model of the grid topology. However, differently from the correct one, the wrong adjacent \acp{ufc} could include non-monitored buses close to the real fault location, which gives the estimation process a degree of freedom that allows to compensate the modelling error. 

To fix this problem, the \acp{se} associated to single-line \acp{ufc} with two monitored buses should be put in condition to work with the same degrees of freedom of those associated to the adjacent \acp{ufc}. The solution is to add \textit{fictitious buses} in the middle of all single-line \acp{ufc} with two monitored buses, and apply the approach to the so obtained extended grid. More formally, we first define this last as \ac{feg}. Then, we can apply Definition~\ref{def:1} to the \ac{feg} to obtain the relevant \acp{ufc}, which, in this case, cannot be single-line. By Theorem~\ref{th:2} there will be a unique \ac{wmr} associated to each of these \acp{ufc}, but now the one corresponding to the right one will return the minimum \ac{wmr}.

Assuming to apply this procedure for single-line \acp{ufc} with two monitored buses, we can define, as follows, the \acp{ufc2}, which coincide with the clusters obtained on the \ac{feg} according to Definition~\ref{def:1}, omitting then the presence of fictitious buses.   

\begin{definition}\label{def:2}
An \ac{ufc2} is a connected portion of the grid, defined as a subset of lines $\mathcal{C}\in\{1,\ldots,m\}$ connected to the rest of the grid through a non-monitored bus with degree $>2$.
\end{definition}

\begin{figure}[t]
  \centering
    \includegraphics[width=1\linewidth]{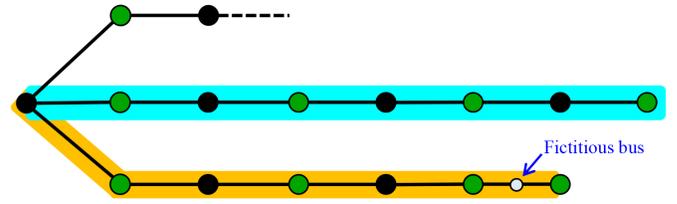}
  \caption{\acp{ufc2} definition applied to the example of Fig.~\ref{fig:clusters}.}
  \label{fig:fake_clusters}
\end{figure}

In Fig.~\ref{fig:fake_clusters}, we can observe how the definition of clusters changes adopting Definition~\ref{def:2} instead of Definition~\ref{def:1}. Notice that \ac{ufc2} are separated by non-monitored fork buses (bus with degree $> 2$).   

It is worth remarking here that all theoretical results introduced in this section do not depend on the dynamic response of all grid components to the fault. Therefore, the \ac{fdcla} and the \acp{pmu} \ac{opa} introduced in the two following sections will work regardless of the presence of dynamic loads and/or distributed generation.

%%%%%%%%%%%%%%%%%%%%%%%%%%%%%%%%%%%%%%%%%%%%%%%%%%%%
\section{Fault Detection and Cluster Localization} \label{sec:fdcla}
%%%%%%%%%%%%%%%%%%%%%%%%%%%%%%%%%%%%%%%%%%%%%%%%%%%%
This section illustrates the \ac{fdcla}, which is able to detect, characterize and locate the occurrence of a fault within one of the grid \acp{ufc2}, using a reduced number of \acp{pmu}, with respect to the \ac{fdla} introduced in Section~\ref{sec:fdla}.

The pseudo-code of the \ac{fdcla} is reported in Algorithm~2. 
First of all, the number and the composition of the grid \acp{ufc2} is determined by function \texttt{\small getUFC}. This function is designed according to Definition~\ref{def:2}, which states that \acp{ufc2} are separated by fork buses. Therefore, using standard graph analysis algorithms, and given the grid topology and the set of monitored buses $\mathcal{M}_b$, \texttt{\small getUFC} returns the number $r<m$ and the composition of the \acp{ufc2}, $\{\mathcal{C}^l\}$, $l=1,2,\ldots,r$, where $\mathcal{C}^l\in\{1,2,\ldots,m\}$. 

Then, at each time step $t$, one state estimate associated to each \acp{ufc2} $x^{l}_t$ and one associated to the grid without virtual buses, the $0$-th $\hat{x}^{0}_t$, and the relevant \acp{wmr} are computed. 

Finally, the same steps of the \ac{fdla} (Algorithm~1) are repeated, simply applying the detection and localization functions \texttt{\small detectFault} and \texttt{\small localizeFault}, exactly as defined in Algorithm~1, to the $r$ estimates associated to the $r$ \acp{ufc2}. In particular, the localization function returns the index $l_F$, that identifies the faulted cluster $\mathcal{C}^{l_F}$. Characterization is realized exactly as in Algorithm~1.

\begin{algorithm}[t] 
 \caption{\ac{fdcla}}
\SetAlgoLined
\DontPrintSemicolon
$[r,\{C^l\}]\leftarrow$  \texttt{getUFC}(\textit{Grid Topology}, $\mathcal{M}_b$)\;
\KwData{$r,\{C^l\},\delta,\mu_{\Delta w^0},th_w,th_{0ng},R,H,\mu_{w}^l, l=1:r$}
\textit{FaultDetected} $\leftarrow$ 0;\;
\textbf{for all} measurement time step $t$ \textbf{do}\;
collect \acp{pmu} measurements $z_t$;\; 
compute $\hat{x}^{l}_t$ and $w^{l}_t$, $\forall l=0,1,\dots,r$;\;
[\textit{FaultDetected}, $t_F$] $\leftarrow$ \texttt{detectFault}($w^0_t$, $w^0_{t-1}$, $\hat{x}_t^l$);\;
\textbf{return} \textit{FaultDetected}, $t_F$;\;
\If{\textit{FaultDetected} \textbf{and} $t=t_F+\delta$}
{$l_F$ $\leftarrow$ \texttt{localizeFault}($\{w^l_{t_F+\tau}\}_{\tau=0,1,\ldots,\delta}$);\;
\textit{FaultedPhases} $\leftarrow$ \texttt{characterizeFault}($\hat{x}_{t_F}^{l_F}$);\;
\textit{FaultDetected} $\leftarrow$ 0;\;
\textbf{return} \textit{FaultedPhases, $\mathcal{C}^{l_F}$};\;}  
\textbf{endfor}\;
\end{algorithm}

%%%%%%%%%%%%%%%%%%%%%%%%%%%%%%%%%%%%%%%%%%%%%%%%%%%%%
\section{Optimal Positioning of \acp{pmu}}
\label{sec:OptimalPositioning}
%%%%%%%%%%%%%%%%%%%%%%%%%%%%%%%%%%%%%%%%%%%%%%%%%%%%%
This section illustrates the \acp{pmu} \acf{opa} by which it is possible to set the number and the positions of \acp{pmu} along with the grid buses, to allow the \ac{fdcla} to work correctly and get a desired fault localization resolution.

\subsection{Localization Resolution}
The localization resolution of the \ac{fdcla} depends on the number $r$ of \acp{ufc2}. According to Definition~\ref{def:2}, $r$ is augmented if fork buses (buses with degree $>2$) are forced to be non-monitored as much as possible. More precisely, it results that
\begin{equation} \label{eq:number_of_ufcs}
    r = 1 + \sum_{\kappa\in\mathcal{N}_{fb}} (\rho(b_\kappa)-1) = 1 + \sum_{\kappa\in\mathcal{N}_{fb}} \rho(b_\kappa) - |\mathcal{N}_{fb}|
\end{equation}
where $\mathcal{N}_{fb}\subseteq \mathcal{N}_{b}$ is the set of non-monitored fork buses. Relation \eqref{eq:number_of_ufcs} follows from the fact that any non-monitored fork bus $b_\kappa\in\mathcal{N}_{fb}$ augments $r$ by  $\rho(b_\kappa)-1$. From \eqref{eq:number_of_ufcs}, it follows that $r$ is maximized by: 
\begin{itemize}
    \item[a)] maximizing the number of non-monitored fork buses $|\mathcal{N}_{fb}|$ (even if it is not apparent in \eqref{eq:number_of_ufcs} we know that $\sum_{\kappa\in\mathcal{N}_{fb}} \rho(b_\kappa) > 2 |\mathcal{N}_{fb}|$ and, thus, $r>1+|\mathcal{N}_{fb}|$), and
    \item[b)] maximizing the degree of non-monitored fork buses.
\end{itemize}
Point b) is important when there are two or more adjacent fork buses. In this case, since we always need to keep observability according to Theorem~\ref{th:1}, not all of these adjacent buses can be non-monitored. Therefore, to augment $r$, we must leave non-monitored fork buses with the lowest degrees.

\subsection{The \acp{pmu}-\ac{opa} algorithm}\label{ssec:opa}
We can now define the \acp{pmu}-\ac{opa}, consisting in the solution of the following mixed-integer optimization problem:   
\begin{align}
	& \min_{\gamma} c^\top \gamma \label{eq:positioningObjective} \\
	& s.t.
    \begin{cases}
            A\gamma\geq{f} \label{eq:positioningConstraints} \\ 
	        \gamma_i=1 &\mbox{if } \rho(b_i)=1 \quad \forall i=1,2,\ldots,n
    \end{cases}
\end{align}
where: $\gamma=[\gamma_1 \ \gamma_2 \ \cdots \gamma_n]^\top$ is a vector of binary variables representing if the $i$-th bus is monitored $(\gamma_i=1)$ or not $(\gamma_i=0)$;
matrix $A\in\mathbb{R}^{n \times n}$ describes the grid topology:
\begin{equation} \label{eq:incidenceMatrix}
    A_{i,k}= 
    \begin{cases}
        \rho(b_i) &\mbox{if } i=k \\
        1 &\mbox{if } (i\neq k) \mbox{ $\wedge$ ($b_i$ and $b_k$ connected)} \\
        0 & \mbox{otherwise}
    \end{cases}
\end{equation}
$f\in\mathbb{R}^n$ with $f_i=\rho(b_i)$; and $c\in\mathbb{R}^n$ collects cost function weights $c_i$.
Constraints in \eqref{eq:positioningConstraints} imposes the grid observability according to Theorem~\ref{th:1}. Cost function weights $c_i$ can be all equal to one if the unique objective is to minimize the number $d$ of required \acp{pmu}, or defined as follows, if the objective is also to maximize the number of \acp{ufc2}:
\begin{equation} \label{eq:cost_function_option2}
    c_i =  \begin{cases}
        1 &\mbox{if } \rho(b_i)\leq 2 \\
        n\cdot \rho(b_i)  & \mbox{if } \rho(b_i)> 2 \\
    \end{cases}.
\end{equation}
Indeed, by applying \eqref{eq:cost_function_option2}, optimization will strongly penalize the positioning of \acp{pmu} on fork buses. If two or more fork buses are adjacent, \eqref{eq:cost_function_option2} will leave non-monitored the one with the highest degree, obtaining the highest possible number of \acp{ufc2}, according to relation \eqref{eq:number_of_ufcs}. 

If the user desires to split of two specific portions of the grid separated by a given bus $b_\kappa$, two scenarios can occur: $b_\kappa$ is a fork bus; or $b_\kappa$ is not a fork bus. 
In the first case, the desired grid partitioning can be obtained applying \eqref{eq:cost_function_option2} if $b_\kappa$ is not adjacent to other fork buses. If the latter occurs, we can set to 1 the weights of fork buses adjacent to $b_\kappa$.

In the second case, we can use the following procedure: 1) extend the grid adding a line, with arbitrary parameters, starting from bus $b_\kappa$ and ending to a fictitious terminal bus $b_{\kappa'}$; 2) set  $\gamma_\kappa=0$ and execute the \ac{opa}, forcing, in this way, the algorithm to put a \ac{pmu} to the terminal fictitious bus $b_{\kappa'}$ and to leave $b_{\kappa}$ non-monitored; 3) install a \ac{pmu} that measures only voltage at bus $b_{\kappa}$; 4) apply the \ac{fdcla} at the grid extended with the fictitious terminal bus $b_{\kappa'}$ assuming that it is monitored but using the voltage measured at bus $b_\kappa$ and setting to zero the measures of the injected currents, with a close to perfect precision (i.e. setting the relevant elements of covariance matrix $R$ to an arbitrary small value $\varepsilon>0$).

\begin{figure}[t]
  \centering
    \includegraphics[width=1\linewidth]{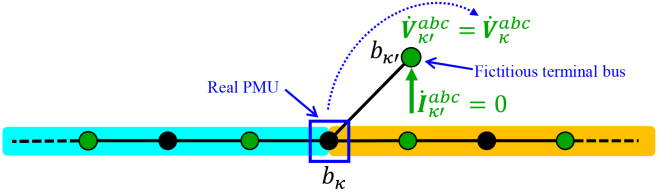}
  \caption{Division of \acp{ufc2} by adding a fictitious terminal bus.}
  \label{fig:cluster-division}
\end{figure}

As we can observe in Fig.~\ref{fig:cluster-division}, this procedure transforms $b_\kappa$ into a fork bus. To set to zero the injected currents at the fictitious terminal bus $b_{\kappa'}$ will lead all \acp{se} to impose that the voltage at the real bus $b_{\kappa}$ is equal to the one of $b_{\kappa'}$, obtaining, in this way, an equivalent and correct representation of the grid. It is worth remarking that the additional \ac{pmu} we need should measure only voltage.  

\subsection{Minimum number of required \acp{pmu}}
The minimum number $d^*$ of required \acp{pmu} depends on the grid topology and on the settings of the \ac{opa} established by the user. Specifically, $d^*$ is influenced by:
\begin{itemize}
    \item $d_1$, defined as the sum of the additive degrees, with respect to 2, of all fork buses, \textit{i.e.}
\begin{equation}
    d_1 = \sum_{\kappa \in \mathcal{I}_{fb}} \left( \rho(b_\kappa)-2 \right) = \sum_{\kappa \in \mathcal{I}_{fb}} \rho(b_\kappa) - 2\cdot |\mathcal{I}_{fb}|,
\end{equation}
where $\mathcal{I}_{fb}\in\{1,2,\ldots,n\}$ is the set of fork buses;
\item $d_2$, equal to the number of fork buses forced to be non-monitored by applying \eqref{eq:cost_function_option2};
\item $d_3$, equal to the number of \acp{pmu} measuring only voltage added to split two portions of the grid separated by a non-fork bus $b_\kappa$, using the procedure above described (Fig.~\ref{fig:cluster-division}).  
\end{itemize}

The influence of these parameters on $d^*$ is provided in the following theorem.
\begin{theorem}\label{th:3}
Consider a radial grid with $n$ buses and $m$ lines. If the \acp{pmu}-\ac{opa} \eqref{eq:positioningObjective}--\eqref{eq:cost_function_option2} is applied with $\gamma^*$ being the solution, than the minimum number of required \acp{pmu} $d^*=\sum_i \gamma_i^*$ is such that
\begin{equation}\label{eq:d_upperbound}
    d^*\leq \overline{d}^* = \lfloor {n}/2 \rfloor+1+d_1 + d_2 + d_3,
\end{equation}
where $\lfloor\cdot \rfloor$ is the \textit{floor function}. 
\end{theorem}
\textit{Proof.} If the grid has no fork bus, we have a simple sequence of $n$ buses. In this case, it easy to show that ${d}^*=\lfloor {n}/2 \rfloor+1$. The presence of a fork bus $b_\kappa$, $\kappa\in \mathcal{I}_{fb}$, of degree $\rho(b_\kappa)$ initializes $\rho(b_\kappa)-2$ new sequences of $\tilde{n}$ additive buses, each requiring up to $\lfloor {\tilde{n}}/2 \rfloor + 1$ additive \acp{pmu}. By summing such an effect for all fork buses, we obtain that  $d^* \leq \lfloor {n}/2 \rfloor + 1 + d_1$. 
Moreover, it can be shown that if a fork bus is forced to be non-monitored, this can cause the requirement of one additive \ac{pmu}. Therefore, the upper-bound $\overline{d}^*$ is augmented by $d_2$. Finally, it is obvious that $\overline{d}^*$ is augmented by $d_3$. 
\vspace{10pt} 

\begin{figure}[t]
  \centering
    \includegraphics[width=1\linewidth]{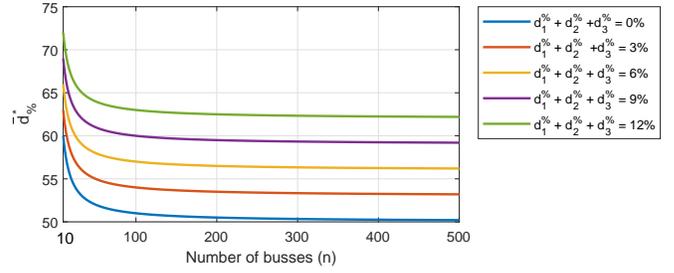}
  \caption{Upper-bounds of the percentage of the minimum number of required \acp{pmu} in function of buses number $n$.}
  \label{fig:required_pmus}
\end{figure}

Theorem~\ref{th:3} provides an upper-bound of the minimum number of required \acp{pmu} in function of the grid topology and of the settings of the \ac{opa} chosen by the user. From \eqref{eq:d_upperbound} we can approximately compute the upper-bound $\overline{d}^*_{\%}$, expressed as the percentage of $\overline{d}^*$ over the total number of buses in the grid: 
\begin{equation} \label{eq:max_pmus_number}
    \overline{d}^*_{\%} \approx  50 \% + 100\%/n +  d_1^{\%} + d_2^{\%} + d_3^{\%} 
\end{equation}
where $d_1^{\%}$, $d_2^{\%}$ and $d_3^{\%}$ are the percentage values of $d_1$, $d_2$, and $d_3$, respectively.
Figure~\ref{fig:required_pmus} shows how $\bar{d}^*_{\%}$ varies in function of the number of grid buses $n$ and of the sum $d_1^{\%}+d_2^{\%}+d_3^{\%}$. Notice that, for $n>250$, $\bar{d}^*_{\%}$ becomes constant with respect to $n$ and it can be approximated as $\bar{d}^*_{\%}\approx 50 \%  +  d_1^{\%} + d_2^{\%} + d_3^{\%}$. 

\subsection{Relation between $\overline{d}^*$ and localization resolution}
Once the \acp{pmu}-\ac{opa} \eqref{eq:positioningObjective}--\eqref{eq:cost_function_option2} is applied, localization resolution $r$ defined in \eqref{eq:number_of_ufcs} becomes:
\begin{equation}\label{eq:ropt}
        r^* = 1 + \sum_{\kappa\in\mathcal{N}_{fb}} \rho(b_\kappa) - |\mathcal{N}_{fb}|  + d_3
\end{equation}

\begin{figure}[t]
  \centering
    \includegraphics[width=1\linewidth]{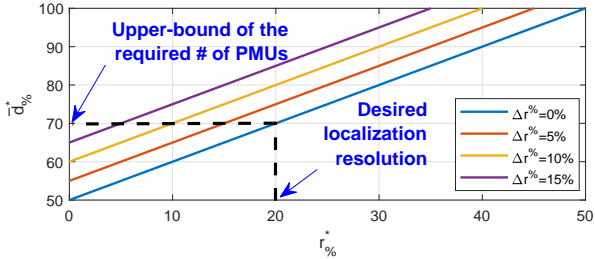}
  \caption{Upper-bounds of required \acp{pmu} $\bar{d}_{\%}^*$ in function of the desired localization resolution $r^*_{\%}$.}
  \label{fig:r-to-d}
\end{figure}

In the particular case of a grid without adjacent fork buses, we have that, applying \eqref{eq:cost_function_option2}, $\mathcal{N}_{fb}=\mathcal{I}_{fb}$ and $d_2 = |\mathcal{I}_{fb}|=|\mathcal{N}_{fb}|$ (all fork buses are forced to be non-monitored). In this case, from \eqref{eq:ropt} and  \eqref{eq:max_pmus_number} we obtain:
\begin{align}
     r^* &= 1 + \sum_{\kappa\in\mathcal{I}_{fb}} \rho(b_\kappa) - |\mathcal{I}_{fb}|  + d_3 \nonumber \\
     &= 1 + d_1 + |\mathcal{I}_{fb}|   + d_3 = 1 + d_1 + d_2  + d_3
\end{align}
from which, taking into account \eqref{eq:d_upperbound}, we have
\begin{equation} \label{eq:relation_r_d}
     r^*= \overline{d}^* - \lfloor {n}/2 \rfloor.
\end{equation}
In the general case, where there are adjacent fork buses, $r^*$ is reduced by the quantity
\begin{equation}
    \Delta r =  \sum_{\kappa\in\mathcal{M}_{fb}} (\rho(b_\kappa)-1) = \sum_{\kappa\in\mathcal{M}_{fb}} \rho(b_\kappa) - |\mathcal{M}_{fb}|
\end{equation}
where $\mathcal{M}_{fb}=\mathcal{I}_{fb}\setminus \mathcal{N}_{fb}$ is the set of monitored fork buses. Indeed, by monitoring a fork bus $b_\kappa$ we lost  $\rho(b_\kappa)-1$ additive \acp{ufc2}. Therefore, the relation between the localization resolution $r^*$ and the relevant upper-bound $\overline{d}^*$ for the required \ac{pmu} is given by:
\begin{equation} \label{eq:relation_r_d_general}
    \overline{d}^*= r^*+ \lfloor {n}/2 \rfloor + \Delta r, 
\end{equation}
which can be expressed in percentage terms as
\begin{equation} \label{eq:relation_r_d_percentage}
     \overline{d}^*_{\%} \approx r^*_{\%}  + 50\% +\Delta r_{\%}, 
\end{equation}
with $\Delta r_{\%}= (\Delta r/n)\cdot 100\%$. 
Figure~\ref{fig:r-to-d} reports relation \eqref{eq:relation_r_d_percentage} for different levels of $\Delta r_{\%}$. 
Notice that the maximal obtainable resolution decreases from $50\%$ realized with $\bar{d}^*_{\%}=100\%$ exactly of $\Delta r^{\%}$.

It is worth remarking that $\bar{d}^*_{\%}=100\%$ does not mean that all buses are equipped with a standard \ac{pmu}, but there are $d_3^{\%}$ buses equipped with a \ac{pmu} measuring only voltage. In any case, it is obvious that, if all buses are monitored by standard \acp{pmu} measuring both voltage and current, we can use the \ac{fdla} introduced in Section~\ref{sec:fdla}, by which we obtain that $r=m=n-1$ and therefore $r_{\%}\approx 100\%$. However, this case is not included in the theory developed in this and the previous section since the presence of just one non-monitored bus triggers the \ac{wmr} clustering phenomenon. In other words, all relations we provided in this section, especially \eqref{eq:relation_r_d_general}-- \eqref{eq:relation_r_d_percentage} are valid only if $d<n$.

%%%%%%%%%%%%%%%%%%%%%%%%
\subsection{Extension to reconfigurable networks}\label{ssec:reconfiguration}
%%%%%%%%%%%%%%%%%%%%%%%%%%%%%%%%%%%%%%%%
All theoretical results and algorithms above introduced are referred to a radial grid with a fixed topology. However, in modern distribution systems, network reconfiguration will become more usual \cite{Su:2019,Dua:2019,Dua:2021}. In this section, we provide simple rules in order to keep the required grid observability and the same fault localization resolution for each possible network topology.

To keep grid observability, the following conditions must be satisfied.
\begin{itemize}
    \item[a.] If buses $b_{\alpha}$ and $b_{\beta}$ are separated by a switch, at least one of them must be monitored (in order to satisfy condition a. of Theorem~\ref{th:1}).
    \item[b.] Let $b_{\alpha}$ be a bus adjacent to a switch. If opening this switch it results that $\rho(b_{\alpha})=1$, then $b_{\alpha}$ must be monitored (in order to satisfy condition b. of Theorem~\ref{th:1}).
\end{itemize}

To satisfy these two conditions, it is enough to add to the \acp{pmu}-\ac{opa}: constraint $\gamma_{\alpha}+\gamma_{\beta}\geq 1$, for condition a. and constraint $\gamma_{\alpha}=1$, for condition b..

To keep the same localization resolution for each possible network configuration, it is sufficient to allow the separation of clusters at switches locations. To obtain this, the following conditions must be satisfied. 
\begin{itemize}
    \item[c.] If a switch is located between two buses $b_{\alpha}$ and $b_{\beta}$ that must be monitored to get fault localizability, at least one of them, say $b_{\alpha}$, must be adjacent to another monitored bus $b_{\kappa}$.
    \item[d.] If opening any switch a bus $b_{\alpha}$, not required to be monitored to get fault localizability, changes its degree from 3 to 2, it must be equipped at least with a \ac{pmu} that measures only voltage. 
\end{itemize}

To satisfy these two conditions it is sufficient to add to the \acp{pmu}-\ac{opa}: constraint $\gamma_{\kappa}=1$, for condition c. and constraint $\gamma_{\alpha}=1$, for condition d..

Condition c. allows to use $b_{\alpha}$ as a separation bus (according to the procedure described in Fig.~\ref{fig:cluster-division}) when the switch is closed (see the example in Section~\ref{sec:results}). Condition d. is referred to the case of a bus that, when the switch is closed, is a fork which separates clusters if non-monitored, but it is no more a fork when the switch is open. In this second situation, the bus does not separate clusters causing a change of the localization resolution. Therefore, condition d. assures that, when the switch is closed, the bus can continue to separate clusters, always using the procedure described in Fig.~\ref{fig:cluster-division}. 

%%%%%%%%%%%%%%%%%%%%%%%%%%%%%%%%%%%%%%%%%%%%%%%%%%%%%
\section{Simulation Results}\label{sec:results}
\begin{figure}[t]
  \centering
    \includegraphics[width=0.7\linewidth]{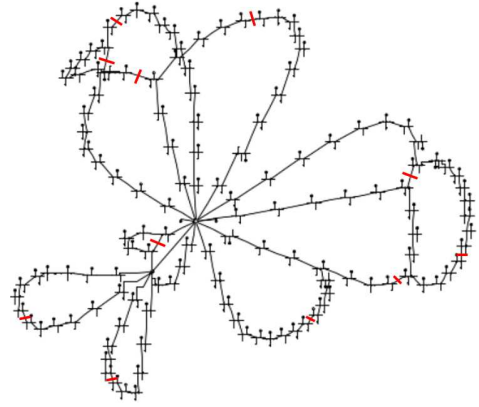}
  \caption{Study case network (1-MV-urban--0-sw in \cite{Meinecke:2020b,Meinecke:2020}).}
  \label{fig:network}
\end{figure}

%%%%%%%%%%%%%%%%%%%%%%%%%%%%%%%%%%%%%%%%%%%%%%%%%%%%%
Figure~\ref{fig:network} shows the network used to test the proposed approach. The study case is a 10 kV distribution grid composed of 142 buses and 147 lines. Details of the network model, implemented in DigSilent PowerFactory, can be found in \cite{Meinecke:2020b,Meinecke:2020}. Red markers in Fig.~\ref{fig:network} indicates normally open switches so that grid is always in radial configuration. Each bus has a load and a PV plant that implements a grid-following inverter-driven generation. Therefore, each PV unit delivers constant power limited by the maximum current value in the case of fault.

\begin{figure}[t]
  \centering
    \includegraphics[width=0.75\linewidth]{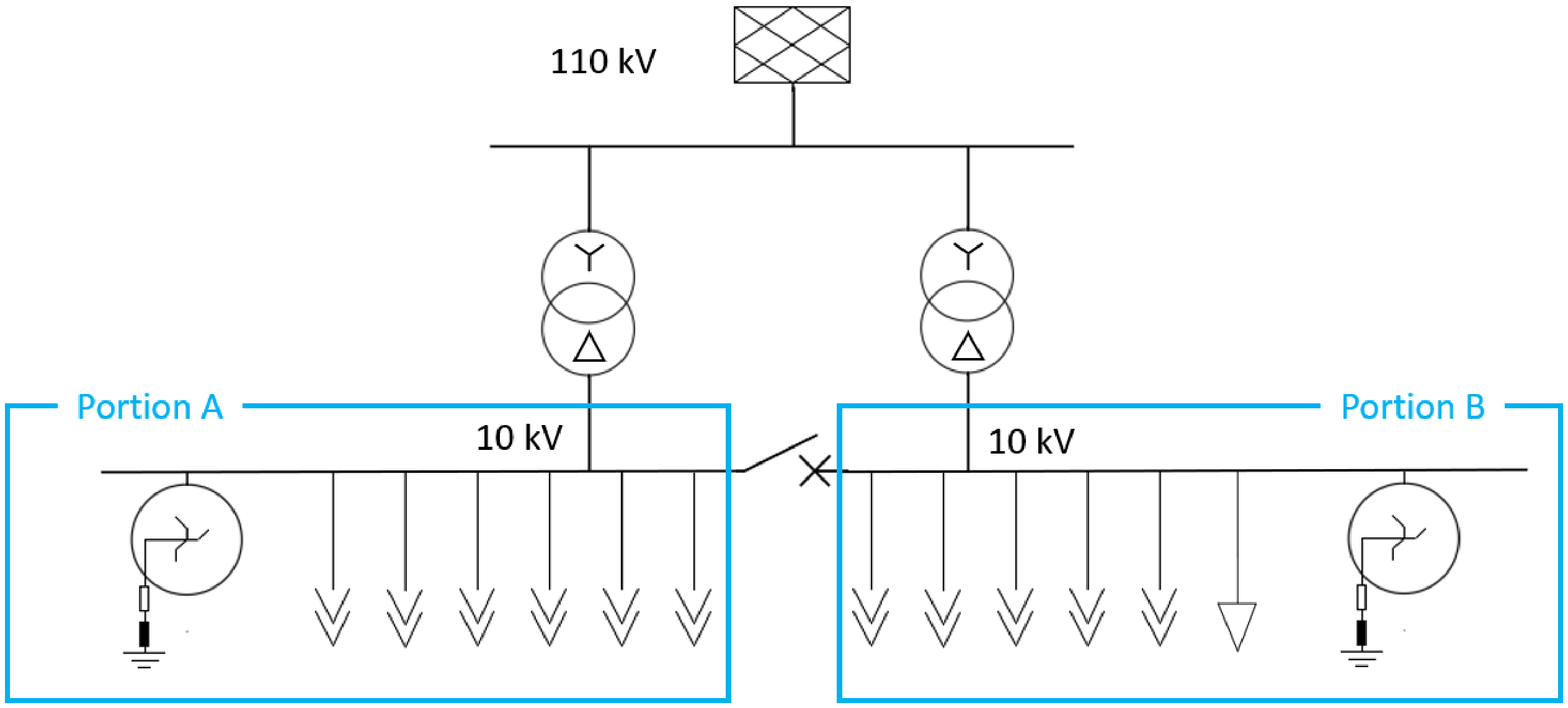}
  \caption{HV/MV interface.}
  \label{fig:network_N}
\end{figure}

As shown in Fig.~\ref{fig:network_N}, the connection with the 110 kV external grid is realized by two HV/MV transformers supplying the two main busbars connected by a normally open tie-breaker. Thus, the network can be divided in two independent portions. In this analysis, we consider only Portion A, composed by 84 buses and 83 lines. The nominal power of the HV/MV transformer of this portion is 63 MVA, whereas the aggregated nominal power of PV generators is 10.6 MVA. Since two grounding transformers make the neutral accessible, two configurations are considered: one with neutral earthed by zero impedance (earthed neutral) and one with neutral earthed by Peterson coil (compensated neutral).

\begin{table*}[t]
		\centering
		\caption{Results out of 100 simulations per scenario.}
		\label{tab:MCResults}
		\renewcommand{\arraystretch}{1}
		\setlength{\tabcolsep}{1.5pt}
		\begin{tabular}{|l| c| c| c| c |c |c| c| c| c| c| c| c| c| c| c| c| c| c| c| c|}
		 \hline
		  \textbf{Fault} & {3ph} &{2ph} &{3ph} &{1ph-e} & {1ph-c} &{2ph} &{3ph} &{1ph-e} & {1ph-c} &{2ph} &{2ph} &{1ph-e} & {1ph-c} &{3ph} &{2ph} &{2ph} & {1ph-e} &{1ph-c} &{1ph-e} &{1ph-c} \\
		  \textbf{Line} & {38-39} &{49-50} &{68-69} &{75-76} & {75-76} &{79-80} &{89-90} &{101-102} & {101-102} &{120-121} &{43-44} &{127-128} & {127-128} &{92-93} &{98-99} &{115-116} & {38-39} &{38-39} &{49-50} &{49-50} \\
		  \textbf{Distance} & {25\%} &{50\%} &{75\%} &{25\%} & {25\%} &{50\%} &{75\%} &{25\%} & {25\%} &{50\%} &{50\%} &{75\%} & {75\%} &{50\%} &{50\%} &{50\%} & {25\%} &{25\%} &{50\%} &{50\%} \\
		  \hline
		  \multicolumn{21}{|c|}{Normal topology}\\
		  \hline
		  \textbf{D-L} &$100$ &$100$ &$100$ &$100$ &$100$ &$100$ &$100$ &$100$ &$100$ &$100$ &$100$ &$100$ &$100$ &$100$ &$100$ &$100$ &$100$ &$34$ &$100$ &$47$ \\ 
		  \textbf{D-nL} &$0$ &$0$ &$0$ &$0$ &$0$ &$0$ &$0$ &$0$ &$0$ &$0$ &$0$ &$0$ &$0$ &$0$ &$0$ &$0$ &$0$ &$66$ &$0$ &$53$ \\ 
		  \hline
		  \multicolumn{21}{|c|}{{Second topology}}\\
		  \hline
		  {\textbf{D-L}} &{$100$} &{$100$} &{$100$} &{$100$} &{$100$} &{$100$} &{$100$} &{$100$} &{$100$} &{$100$} &{$100$} &{$100$} &{$100$} &{$100$} &{$100$} &{$100$} &{$100$} &{$30$} &{$100$} &{$52$} \\ 
		  {\textbf{D-nL}} &{{$0$}} &{$0$} &{$0$} &{$0$} &{$0$} &{$0$} &{$0$} &{$0$} &{$0$} &{$0$} &{$0$} &{$0$} &{$0$} &{$0$} &{$0$} &{$0$} &{$0$} &{$70$} &{$0$} &{$48$} \\ 
		  \hline
\end{tabular}
%\vspace{-10pt}
\end{table*}

\begin{figure}[t]
  \centering
    \includegraphics[width=\linewidth]{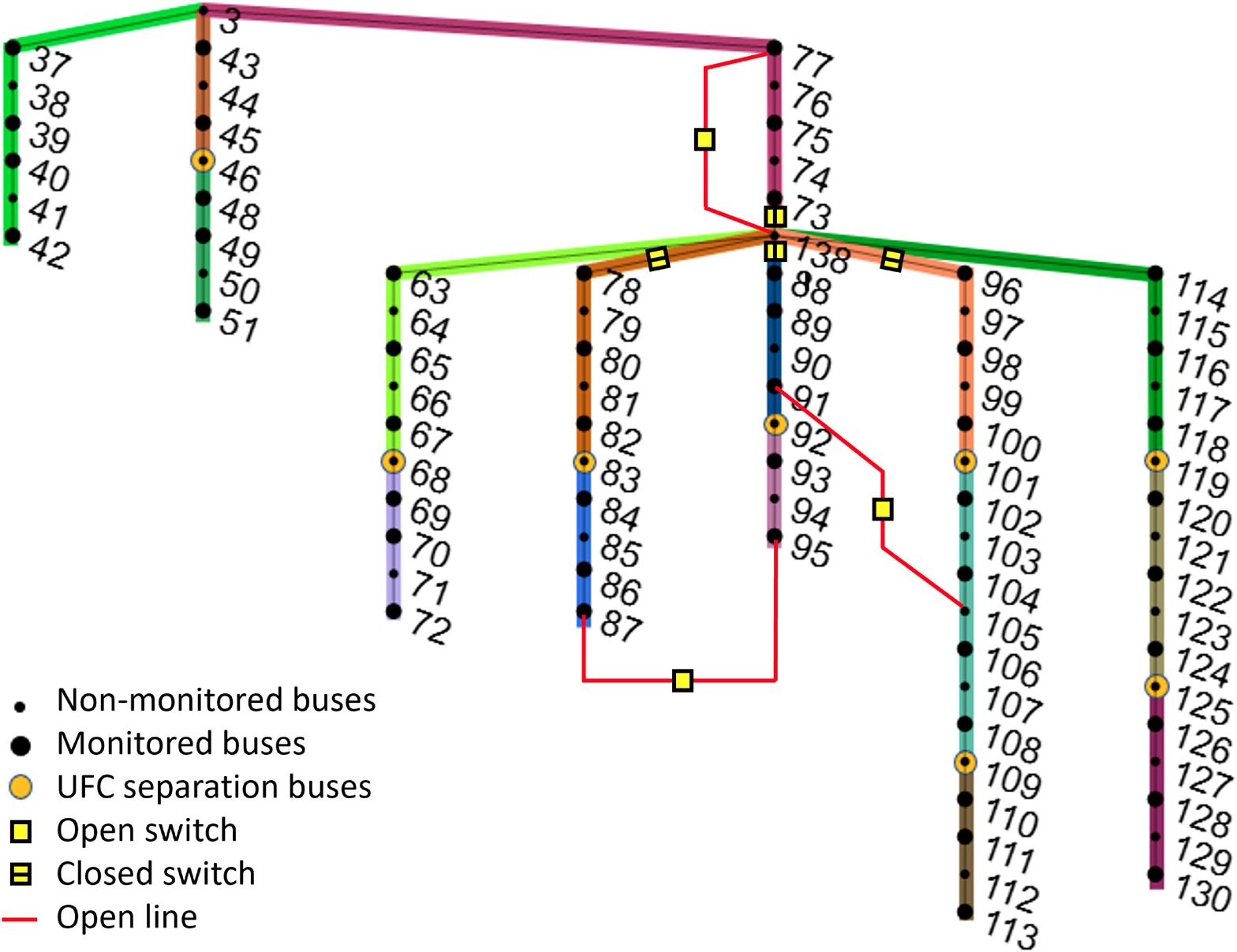}
  \caption{Normal network topology with the position of monitored and non-monitored buses and \acp{ufc2} indicated by different colours.}
  \label{fig:graph}
\end{figure}
\begin{figure}[t]
  \centering
    \includegraphics[width=1\linewidth]{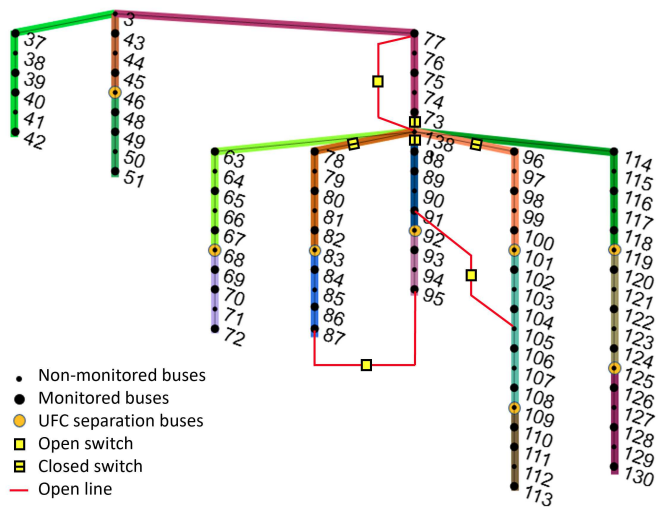}
  \caption{{Second network topology with the position of monitored and non-monitored buses and \acp{ufc2} indicated by different colours.}}
  \label{fig:graph2}
\end{figure}

The normal topology of the considered grid is represented in Fig.~\ref{fig:graph}, where monitored buses are bold black dots and non-monitored buses are small black dots. Figure~\ref{fig:graph} also reports switches and normally open lines (red lines). Therefore, it is assumed that the grid configuration can be changed by suitably opening and closing switches, always keeping a radial topology.

\acp{pmu} are placed using the \acp{pmu}--\ac{opa} introduced in Section~\ref{ssec:opa}, {taking into account the conditions introduced in Section~\ref{ssec:reconfiguration} to keep fault localizability and localization resolution in case of network reconfiguration}. As expected, fork buses 3 and 138 are forced to be non-monitored by the optimization problem \eqref{eq:positioningObjective}--\eqref{eq:cost_function_option2}, so that the grid is divided into 8 \acp{ufc2}. The number of required \acp{pmu} results to be $48$. The number of clusters is then increased to $16$ by adding 8 voltage meters at non-monitored buses, marked by a yellow circle in Fig.~\ref{fig:graph}, according to the procedure explained in Fig.~\ref{fig:cluster-division}. The final total number of monitored buses results to be $d=56$. 

We can verify that \eqref{eq:d_upperbound} is satisfied. Indeed, $d_1=5$ (the degrees of fork buses are 3 and 6), $d_2=2$ and $d_3=8$, and, therefore, $\overline{d}^*=58>d$. Moreover, we can verify \eqref{eq:relation_r_d_general} by observing that the resolution returned by the \acp{pmu}--\ac{opa} is $r^*=8$, the incremented resolution is $\Delta r=8$.  

Figure~\ref{fig:graph2} shows a second network topology considered in the simulations: the one obtained by closing the line between buses 87 and 85 and opening the switches between lines 78 and 138. We can observe how bus 87 is transformed into a separation bus (yellow circle) and the composition of the clusters is preserved, except to the fact that line 78--138 becomes an open line, whereas the normally open line 87--95 is included in the \ac{ufc2} between buses 92 and 87.   

\begin{figure*}
     \centering
     \begin{subfigure}[b]{0.32\textwidth}
         \centering
         \includegraphics[width=\textwidth]{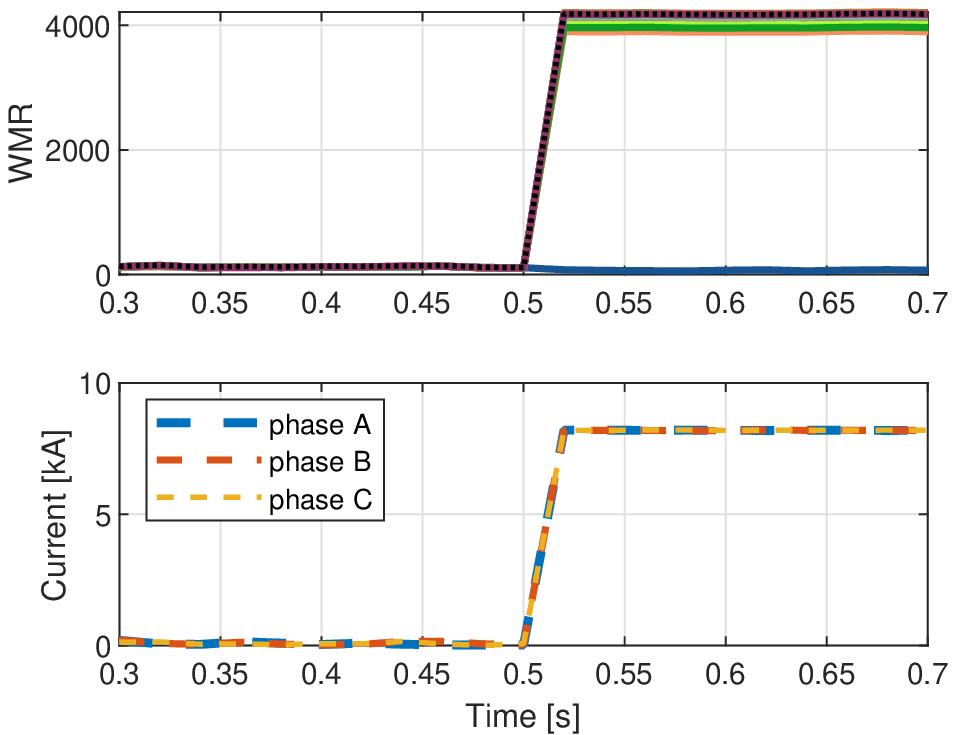}
         \caption{3ph fault at 75\% of line 89-90}
         \label{fig:three-phase fault}
     \end{subfigure}
     \hfill
     \begin{subfigure}[b]{0.32\textwidth}
         \centering
         \includegraphics[width=\textwidth]{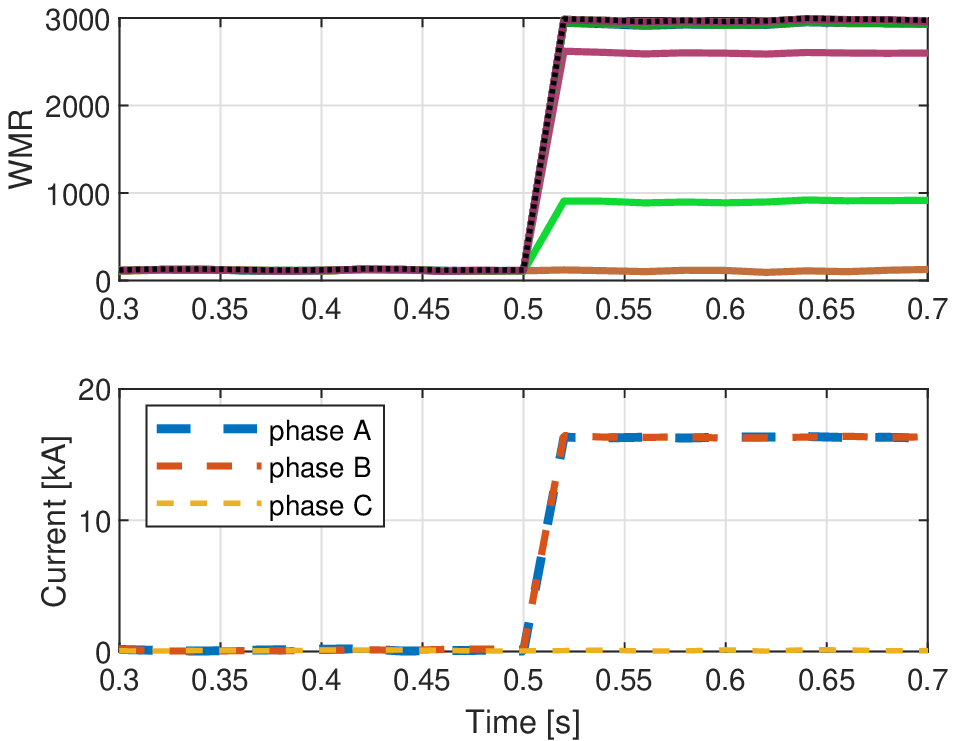}
         \caption{2ph fault at 25\% of line 43-44}
         \label{fig:two-phase fault}
     \end{subfigure}
     \hfill
     \begin{subfigure}[b]{0.32\textwidth}
         \centering
         \includegraphics[width=\textwidth]{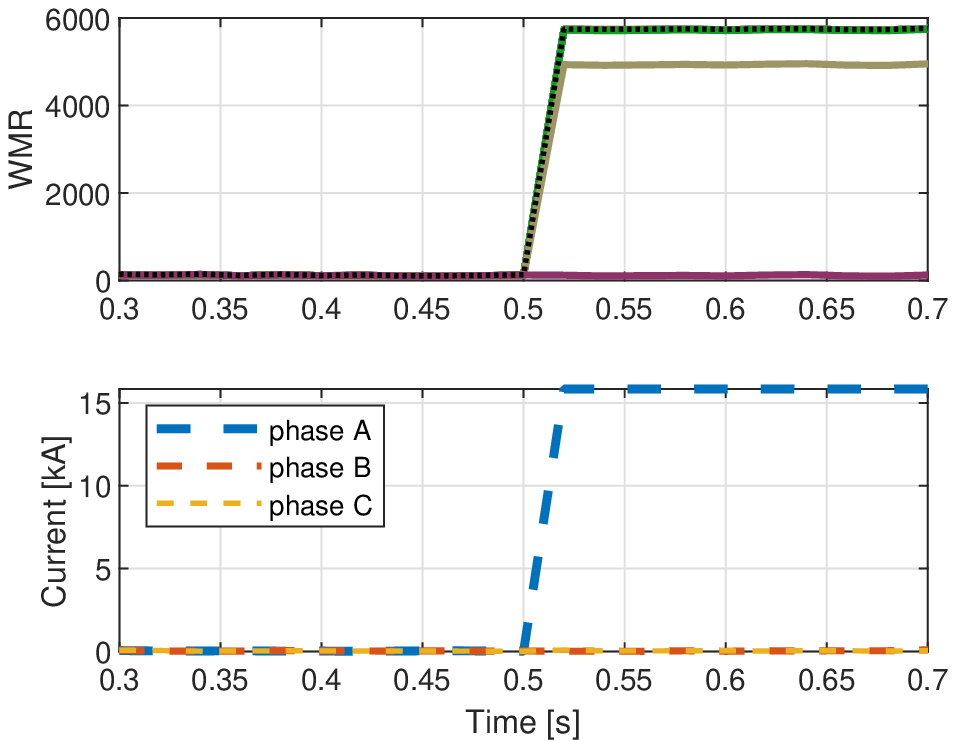}
         \caption{1ph-c fault at 75\% of line 127-128}
         \label{fig:one-phase-g fault}
     \end{subfigure}
        \caption{\acp{wmr} and estimated currents at the virtual bus in three sample cases with normal network topology.}
        \label{fig:wmr-faultcurrents}
\end{figure*}

Measurements are simulated by adding noise to magnitude and phase of voltages and currents. Noises are assumed to be white and Gaussian, with the following standard deviations: $1.6\cdot10^{-3}\%$ and $4\cdot10^{-1}\%$ for voltage and current magnitudes, respectively; $5.1 \cdot 10^{-5}\si{\rad}$ and $5.8 \cdot 10^{-3}\si{\rad}$ for voltage and current phases, respectively. Measures are sampled with a time step $\Delta t$=\SI{20}{\milli \second}. Since measurements are in polar coordinates, a transformation to rectangular coordinates is applied \cite{Milano:2016}. All these assumptions and noise standard deviations are realistic, as discussed in \cite{Pignati:2016}.

Table \ref{tab:MCResults} reports the results obtained by the \ac{fdcla} in a Monte Carlo analysis, with 100 noise realizations, in 20 different fault scenarios, {first with the normal and then with the second topology.} The following types of faults are simulated: 3-phase-to-ground (3ph), phase-to-phase (2ph), phase-to-ground with earthed neutral (1ph-e) and with compensated neutral (1ph-c). The first row of the table reports the fault scenario, while the others collect the number of times in which the fault has been a) D-L: detected and correctly localized, b) D-nL: detected, but localized in the wrong cluster.

First observe that the algorithm successfully detects the fault in all the simulated cases. In most of the scenarios, the algorithm also perfectly locates the fault. In a couple of single-phase fault scenarios with compensated neutral, localization results are not satisfactory. However, consider that the \ac{fdcla} has been tested with $\delta=0$, \textit{i.e.} using just one measurement after the fault time. As explained in Section~\ref{ssec:localization}, localization can be enhanced using more measurements,  
accepting a time delay $\delta>0$. This is verified in Table~\ref{tab:MCResults2}, which shows the results obtained by progressively augmenting $\delta$ in the two critical fault scenarios. 

\begin{table}[t]
		\centering
		\caption{Successful localization results out of 100 simulations per scenario with $\delta>0$.}
		\label{tab:MCResults2}
		\renewcommand{\arraystretch}{1}
		\begin{tabular}{l c c c c}
		\hline \hline	\vspace{-8pt} \\
		  \textbf{Fault} & {$\mathbf{\delta}$ = 2} &{$\mathbf{\delta}$ = 3} &{$\mathbf{\delta}$ = 4} &{$\mathbf{\delta}$ = 5} \\  \hline \vspace{-5pt}\\
          {\textbf{Normal topology}} & & & & \\
		  {1ph-c at $25\%$ of Line 38-39} &$59$ &$72$ &$75$ &$88$ \\
		  {1ph-c at $50\%$ of Line 49-50} &$88$ &$94$ &$96$ &$99$ \\ \vspace{-5pt}\\
		  {\textbf{Second topology}} & & & & \\
		  {1ph-c at $25\%$ of Line 38-39} &{$64$} &{$74$} &{$75$} &{$82$} \\
		  {1ph-c at $50\%$ of Line 49-50} &{$86$} &{$92$} &{$94$} &{$96$} \\
		  \hline\hline
\end{tabular}
\end{table}

Figure~\ref{fig:wmr-faultcurrents} shows how the \acp{wmr} and the currents estimated at the virtual bus evolve in the case of a 3ph, a 2ph and a 1ph-c fault occurring at $0.5\si{\second}$, with the normal network topology. 
\acp{wmr} trajectories are plotted with the same colors used in Fig.~\ref{fig:graph} to associated them with the corresponding \acp{ufc2}. Many \acp{wmr} trajectories are not visible since they are almost coincident each other. Black dashed lines refer to the $0$-th \ac{wmr} ($w^0_t$), used for fault detection. We can observe how, in all the three examples, $w^0_t$ significantly increases at the fault time allowing a correct detection using \eqref{eq:detection_rule_wmr}. Moreover, there is always a unique \ac{wmr} trajectory that does not increase at the fault time, allowing a correct localization using \eqref{eq:localization_rule}. Finally, in the bottom figures, we can note how the estimated fault currents allow a correct characterization, since only the correct phase currents increase. 

\begin{figure}[t]
  \centering
    \includegraphics[width=1\linewidth]{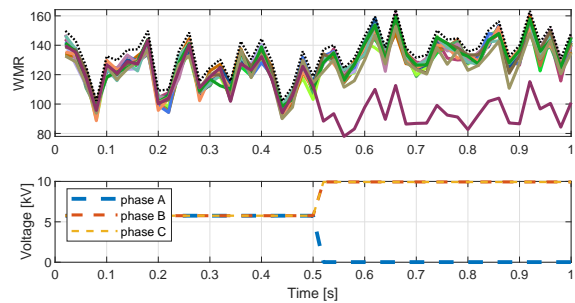}
  \caption{1ph-c fault at 75\% of line 127-128 with normal network topology: \acp{wmr} and 3-phase nodal voltage at the virtual bus.}
  \label{fig:one-phase-p fault}
\end{figure}

Figure~\ref{fig:one-phase-p fault} shows what happens in the particular case of a single-phase fault with compensated neutral. $w_t^0$ does not significantly increase at the fault time. Therefore, \eqref{eq:detection_rule_wmr} does not allow the fault detection, which is however realized using the estimated zero-sequence injected currents, according to rule \eqref{eq:detection_rule_zero_sequence}. 

Moreover, the \ac{wmr} referred to the cluster where the event has occurred is not evidently different from the others as in the cases shown in Fig.~\ref{fig:wmr-faultcurrents}, especially after a single time step from the fault time. This explains why localization returned by \eqref{eq:localization_rule} is not always correct. However, in the figure we clearly observe that there is a difference among the right \ac{wmr} trajectory and the others. As proved by results in Table~\ref{tab:MCResults2}, this can be correctly detected by comparing the average values before and after the fault time, using $\delta>0$ in \eqref{eq:localization_rule_delay}.
Since, because of compensation, fault currents are low, characterization cannot be realized using the estimated fault currents. Therefore, as explained in Section~\ref{ssec:characterization}, the fault is characterized by observing the three-phase nodal voltages at the virtual bus, reported in Fig.~\ref{fig:one-phase-p fault}.

%%%%%%%%%%%%%%%%%%%%%%%%%%%%%%%%%%%%%%%%%%%%%%%%%%%%%
\section{Conclusions}\label{sec:conclusions}
%%%%%%%%%%%%%%%%%%%%%%%%%%%%%%%%%%%%%%%%%%%%%%%%%%%%%
A fault detection and localization algorithm for active distribution networks has been developed in this paper along with an analysis of the required number of \acp{pmu} and their optimal positioning, {considering also the case of reconfigurable networks.}
The approach has been tested by a Monte Carlo analysis on a benchmark MV active distribution network. The algorithm always succeeds in detecting and localizing the fault with the exception of single-phase-to-ground faults, where localization is perfectly achieved only in the case of earthed neutral configuration.   
Single-phase-to-ground faults with compensated neutral are generally more difficult to localize due to the smaller fault current. However, in this case, the algorithm performance are improved by using a set of measurements collected after the fault occurrence, accepting a delay in the localization.

\appendix[Proof of Theorem~\ref{th:2}]

\begin{figure}[t]
 \centering
    \includegraphics[width=1\linewidth]{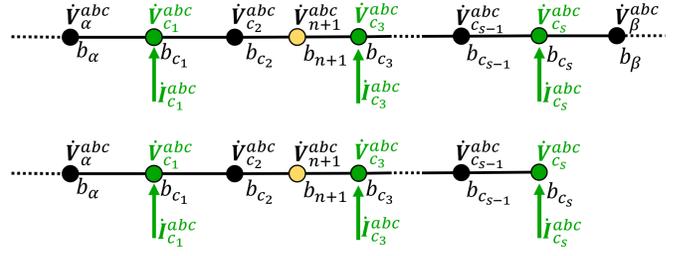}
 \caption{Examples of multi-lines \acp{ufc}. Top: non-terminal. Bottom: terminal.  Green buses are monitored; black buses are not monitored; the virtual bus $b_{n+1}$ is yellow and not monitored. Green variables are measured.}
 \label{fig:proof_ufc}
\end{figure}

Theorem~\ref{th:2} is trivially proved for single-line \acp{ufc}. Therefore, we focus on multi-lines \acp{ufc}. As detailed in Fig.~\ref{fig:proof_ufc}, in the non-terminal case, a multi-line \ac{ufc} is connected to the rest of the grid through two non-monitored buses $b_{\alpha}$ and $b_{\beta}$. Between these two border buses there are $s+1$ lines connecting $s$ buses, with $s$ being always an odd number. In the terminal case, a multi-lines \ac{ufc} is connected to the rest of the grid through one non-monitored bus $b_{\alpha}$ and it is composed by $s$ lines connecting $s$ buses, with $s$ being again an odd number. In both the cases, $(s+1)/2$ buses are monitored and $(s-1)/2$ are non-monitored.

Supposing that the virtual bus $b_{n+1}$ is placed in the middle of any line $k\in\mathcal{C}$, we define the portion of the extended state $x^k$ corresponding to the $s$ non-border buses $b_{c_l}$, $l = 1,2,\ldots,s$ and to the virtual bus, as follows:
\begin{equation}
\begin{aligned}
    &x^k_{\mathcal{C}}= \\
    &\begin{bmatrix}   \mathbf{V}^{abc}_{c_1,re} \ \cdots  \ \mathbf{V}^{abc}_{c_s,re} \ \mathbf{V}^{abc}_{n+1,re} \ \mathbf{V}^{abc}_{c_1,im}\ \cdots \ \mathbf{V}^{abc}_{c_s,im}\ \mathbf{V}^{abc}_{n+1,im}\end{bmatrix}^\top \label{eq:x_proof}
\end{aligned}
\end{equation}
Notice that $x^{k}_{\mathcal{C}}\in\mathbb{R}^q$ with $q=6\cdot(s+1)$. Moreover, we indicate with $x^k_\alpha$ and $x^k_\beta$ the portions of $x^k$ corresponding to the two border buses $b_\alpha$ and $b_\beta$:
\begin{equation}
\begin{aligned}
    x^k_{\alpha} = \begin{bmatrix}   \mathbf{V}^{abc}_{\alpha,re} & \mathbf{V}^{abc}_{\alpha,im}\end{bmatrix}^\top, \quad
        x^k_{\beta} = \begin{bmatrix}   \mathbf{V}^{abc}_{\beta,re} & \mathbf{V}^{abc}_{\beta,im}\end{bmatrix}^\top.
\end{aligned}
\end{equation}
Finally, we indicate with $\overline{x}^{k}_{\mathcal{C}}$ the rest of the state vector $x^k$, which can be re-ordered as follows:
\begin{equation}
x^k = 
    \begin{bmatrix}
    {x^{k}_{\mathcal{C}}}^\top & {x^{k}_{\alpha}}^\top & {x^{k}_{\beta}}^\top & {\overline{x}^k_{\mathcal{C}}}^\top
    \end{bmatrix}^\top.
\end{equation}
Note that if $\mathcal{C}$ is terminal, $x_{\beta}$ will belong to $\overline{x}^k_{\mathcal{C}}$.  

In a similar manner, we can re-order the output vector $z$ as
\begin{equation}
\label{eq:z_reordered}
    z = \begin{bmatrix}
    {z_{\mathcal{C}}}^\top &
    {\overline{z}_{\mathcal{C}}}^\top
    \end{bmatrix}^\top,
\end{equation}
where $z_{\mathcal{C}}$ is composed by the voltages and injected currents measured at the $(s+1)/2$ monitored buses connected by the lines in $\mathcal{C}$ (\textit{i.e.}, according to Fig.~\ref{fig:proof_ufc}, buses $b_{c_1},b_{c_3},\ldots,b_{c_s}$) and where $\overline{z}_{\mathcal{C}}$ are the measurements collected from buses outside the cluster. Notice that $z_{\mathcal{C}}\in\mathbb{R}^q$, \textit{i.e.} it has the same size of $x^k_{\mathcal{C}}$. Notice also that the latter is not true if we add to the \ac{ufc} a line starting from one of the non-border non-monitored buses (violating condition b.3 of Theorem~\ref{th:2}).

According to \eqref{eq:z_reordered}, and under the hypothesis that measurements collected from different \acp{pmu} are independent, the covariance matrix $R$ can be re-ordered to have the following form:
\begin{equation}
    R =\begin{bmatrix}
    R_{\mathcal{C}} & 0 \\ 0 & \overline{R}_{\mathcal{C}},
    \end{bmatrix}
\end{equation}
where $R_{\mathcal{C}}$ and $\overline{R}_{\mathcal{C}}$ are the covariance matrices relevant to $z_{\mathcal{C}}$ and $\overline{z}_{\mathcal{C}}$, respectively.

Given the above introduced definitions, the output equation of the network extended with a virtual bus placed on the $k$-th line, with $k\in{\mathcal{C}}$, can be rewritten as follows:
\begin{equation}
\label{eq:cluster_output_equation}
    \begin{bmatrix}
    z_{\mathcal{C}} \\
    \overline{z}_{\mathcal{C}}
    \end{bmatrix} =
    \begin{bmatrix}
    H^k_{\mathcal{C}} & h^k_\alpha & h^k_\beta & 0\\
    0& \overline{h}_{\alpha} & \overline{h}_{\beta}& \overline{H}_{\mathcal{C}}
    \end{bmatrix}    \begin{bmatrix}
    x^{k}_{\mathcal{C}} \\ x^{k}_{\alpha} \\ x^{k}_{\beta} \\ \overline{x}^k_{\mathcal{C}}
    \end{bmatrix} + v,
\end{equation}
where $H^k_{\mathcal{C}}\in\mathbb{R}^{q\times q}$, $h^{k}_{\alpha}, h^{k}_{\beta}\in \mathbb{R}^{q}$, $\overline{h}_\alpha, \overline{h}_\beta\in\mathbb{R}^{D-q}$, $H_{\mathcal{C}}\in\mathbb{R}^{(D-q)\times (N+4-q)}$ are composed by the elements of $H^k$. If the \ac{ufc} is terminal, the third block column in the output matrix in \eqref{eq:cluster_output_equation} is missing. 

It is important to note that $H^k_{\mathcal{C}}$ is a square matrix and that, whereas $H^k_{\mathcal{C}}$, $h^{k}_{\alpha}$, and $h^{k}_{\beta}$ vary with $k\in\mathcal{C}$ (i.e. they vary as the position of the virtual bus is changed within the cluster), $\overline{H}_{\mathcal{C}}$, $\overline{h}_{\alpha}$, and $\overline{h}_{\beta}$ are the same for all $k\in\mathcal{C}$. This is obvious since a different placement of the virtual bus within the \ac{ufc} does not modify the topology outside from it. 

From \eqref{eq:cluster_output_equation}, omitting noise, we can obtain the following relation among the variables and the measurements involved in the \ac{ufc}:
\begin{equation}
\label{eq:cluster_relation}
    z_{\mathcal{C}} = H^k_{\mathcal{C}} x^k_{\mathcal{C}} + h^k_{\alpha} x^k_{\alpha} + h^k_{\beta} x^k_{\beta}, 
\end{equation}
where last term is missing if the \ac{ufc} is terminal. 

Consider now the general scenarios in Fig.~\ref{fig:proof_ufc}. It can be shown by electrotechnical computations that given the border buses voltages $\dot{\mathbf{V}}_\alpha^{abc}$ and $\dot{\mathbf{V}}_\beta^{abc}$ (only $\dot{\mathbf{V}}_\alpha^{abc}$ in the terminal case), all voltages of the non-border buses $\dot{\mathbf{V}}_{c_l}^{abc}$, $l = 1,2,\ldots,s$ and the voltage of the virtual bus $\dot{\mathbf{V}}_{n+1}^{abc}$ can be \textit{uniquely} computed using the available measurements collected in $z_{\mathcal{C}}$. This holds true for all possible positions of the virtual bus within the \ac{ufc} (for all lines $k\in\mathcal{C}$). 

Such a fact implies that, for all $k\in\mathcal{C}$, matrix $H^k_{\mathcal{C}}$ is nonsingular and relation \eqref{eq:cluster_relation} can be inverted as follows:
\begin{equation}
\label{eq:inverted_cluster_relation}
    x^k_{\mathcal{C}} = {H^k_{\mathcal{C}}}^{-1}(z_{\mathcal{C}}  - h^k_{\alpha} x^k_{\alpha} - h^k_{\beta} x^k_{\beta}). 
\end{equation}

Notice that this conclusion is not valid if we add to the \ac{ufc} a line starting from one of the non-monitored bus, violating condition b.3 of Theorem~\ref{th:2}. Indeed, in this case, matrix $H_{\mathcal{C}}$ would be no more square (measurements become more than variables to be estimated) and the solution of \eqref{eq:cluster_relation} with respect to $x_{\mathcal{C}}^k$ would be not unique.

Consider now the following function:
\begin{align}
        W(x^k) = & (z_{\mathcal{C}}- H^k_{\mathcal{C}} {x}^k_{\mathcal{C}} - h^k_{\alpha} {x}^k_{\alpha} - h^k_{\beta} {x}^k_{\beta})^\top R_\mathcal{C}^{-1}(-) \nonumber \\
        &+(\overline{z}_{\mathcal{C}}- \overline{h}_{\alpha} {x}^k_{\alpha} - \overline{h}_{\beta} {x}^k_{\beta} - \overline{H}_{\mathcal{C}} {\overline{x}}^k_{\mathcal{C}} )^\top \overline{R}_\mathcal{C}^{-1}(-),
\end{align}
where $(-)$ indicates the same term on the left of the inverted covariance matrices.
Using \eqref{eq:x_proof}--\eqref{eq:cluster_output_equation} and \eqref{eq:wmr}, it is easy to show that $w^k=\sqrt{W(\hat{x}^k)}$. 

Recall now that \ac{wls} estimates \eqref{eq:wlsk} are such that $\hat{x}^k = \arg \min_{x^k} W(x^k)$ and, consequently, 
${w^k} = \sqrt{\min_{x^k}{W(x^k)}}$. In virtue of \eqref{eq:inverted_cluster_relation}, for all $k\in\mathcal{C}$, there always exists a unique $\hat{x}^k_{\mathcal{C}}$ such that, whatever given the value of $x_{\alpha}^k$, $x_\beta^k$, and $\zeta_{\mathcal{C}}$,   
\begin{equation}
\label{eq:cluster_relation1}
    z_{\mathcal{C}} - H^k_{\mathcal{C}} \hat{x}^k_{\mathcal{C}} - h^k_{\alpha} x^k_{\alpha} - h^k_{\beta} x^k_{\beta} = 0. 
\end{equation}
This implies that, for all $k\in{\mathcal{C}}$,
\begin{align}
    &(w^k)^2 = \min_{x^k} W(x^k)= \nonumber \\
    &\min_{x^k_{\alpha},x^k_{\beta},\overline{x}^k_{\mathcal{C}}} (\overline{z}_{\mathcal{C}}- \overline{h}_{\alpha} {x}^k_{\alpha} - \overline{h}_{\beta} {x}^k_{\beta} - \overline{H}_{\mathcal{C}} {\overline{x}}^k_{\mathcal{C}} )^\top \overline{R}_\mathcal{C}^{-1}(-) \label{eq:wmr_reduced}
\end{align}
since the minimum of the first term of $W(x^k)$ is always zero, independently of the minimum of the second term. Because all parameters in the function to be minimized in \eqref{eq:wmr_reduced} do not vary with $k\in{\mathcal{C}}$, the minimum does not change with $k$. This finally proves Theorem~\ref{th:2} since it means that for all $i,\ell\in\mathcal{C}$, $w^{i}=w^{\ell}$.

\ifCLASSOPTIONcaptionsoff
  \newpage
\fi

%%%%%%%%%%%%%%%%%%%%%%%%%%%%%%%%%%%%%%%%%%
\bibliographystyle{IEEEtran}
%\bibliography{IEEEabrv,biblio.bib}

%%%%%%%%%%%%%%%%%%%%%%%%%%%% BIOS
\begin{IEEEbiography}[{\includegraphics[width=1in,height=1.25in,clip,keepaspectratio]{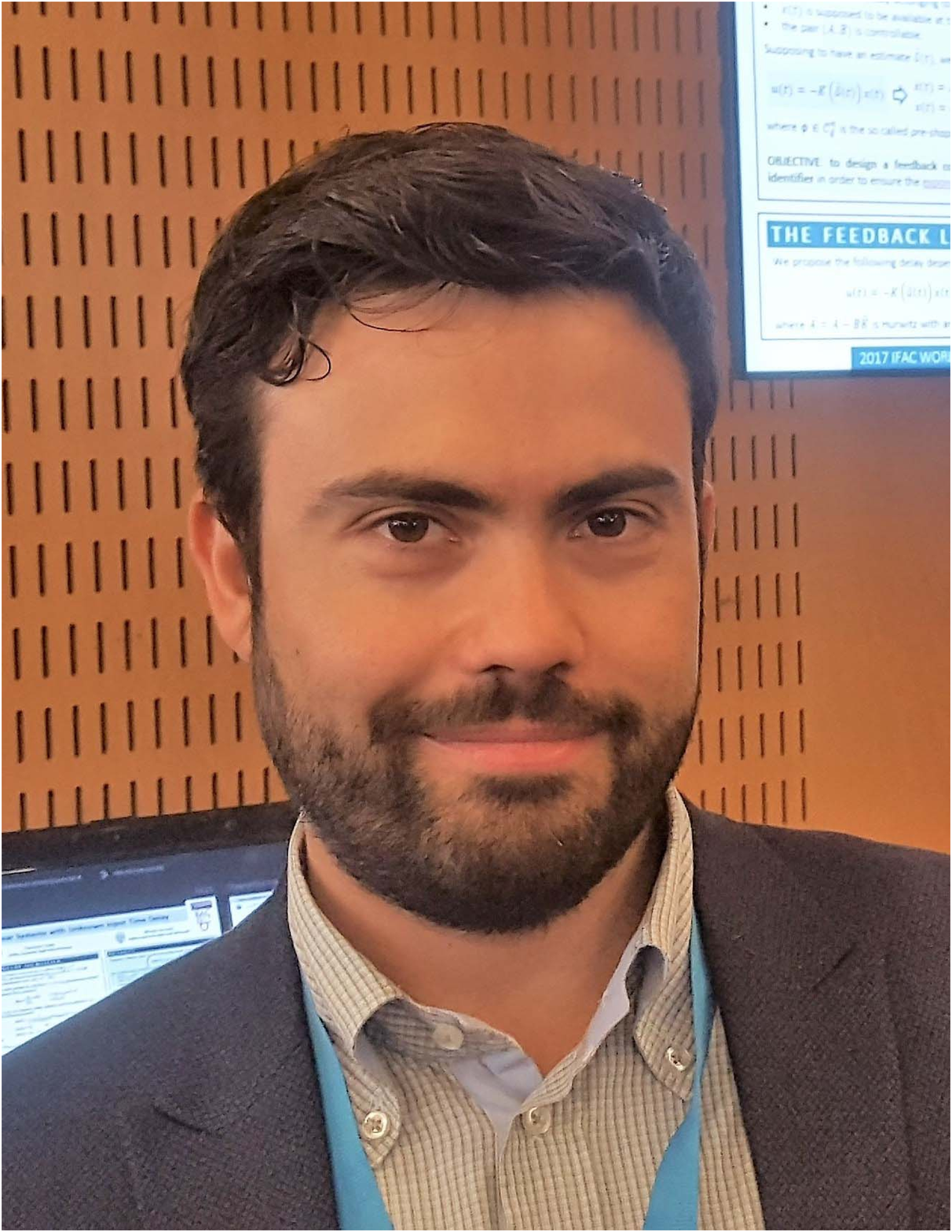}}]{Francesco Conte}(S’11 - M’14 - SM ‘20) received the master’s degree in computer science and automatic engineering and the Ph.D. degree in electrical and information engineering from the University of L’Aquila, Italy, in 2009 and 2013, respectively. 

Since 2021 he is Assistant Professor (tenure track) at the Campus Bio-Medico University of Rome, Italy. From 2012 to 2021 he was with the Department of Electrical, Electronics and Telecommunication Engineering and Naval Architecture, University of Genova, Italy. From December 2008 to April 2009, he was a Visiting Scholar with the French National Institute for Research in Computer Science and Control (INRIA), Grenoble, France. He is the author of more than 60 publications in international journals, books and conferences. His research interests include power system modelling and control, intelligent management of renewable energy sources and loads, distributed generation, smart grids, and local energy communities. 

Dr. Francesco Conte has been principal investigator of five research projects and involved in the working group of two European projects and of other twelve research projects. Since 2018 he is member of the Cigré working group TOR-JWG N° D2/C6.47.
\end{IEEEbiography}

\begin{IEEEbiography}[{\includegraphics[width=1in,height=1.25in,clip,keepaspectratio]{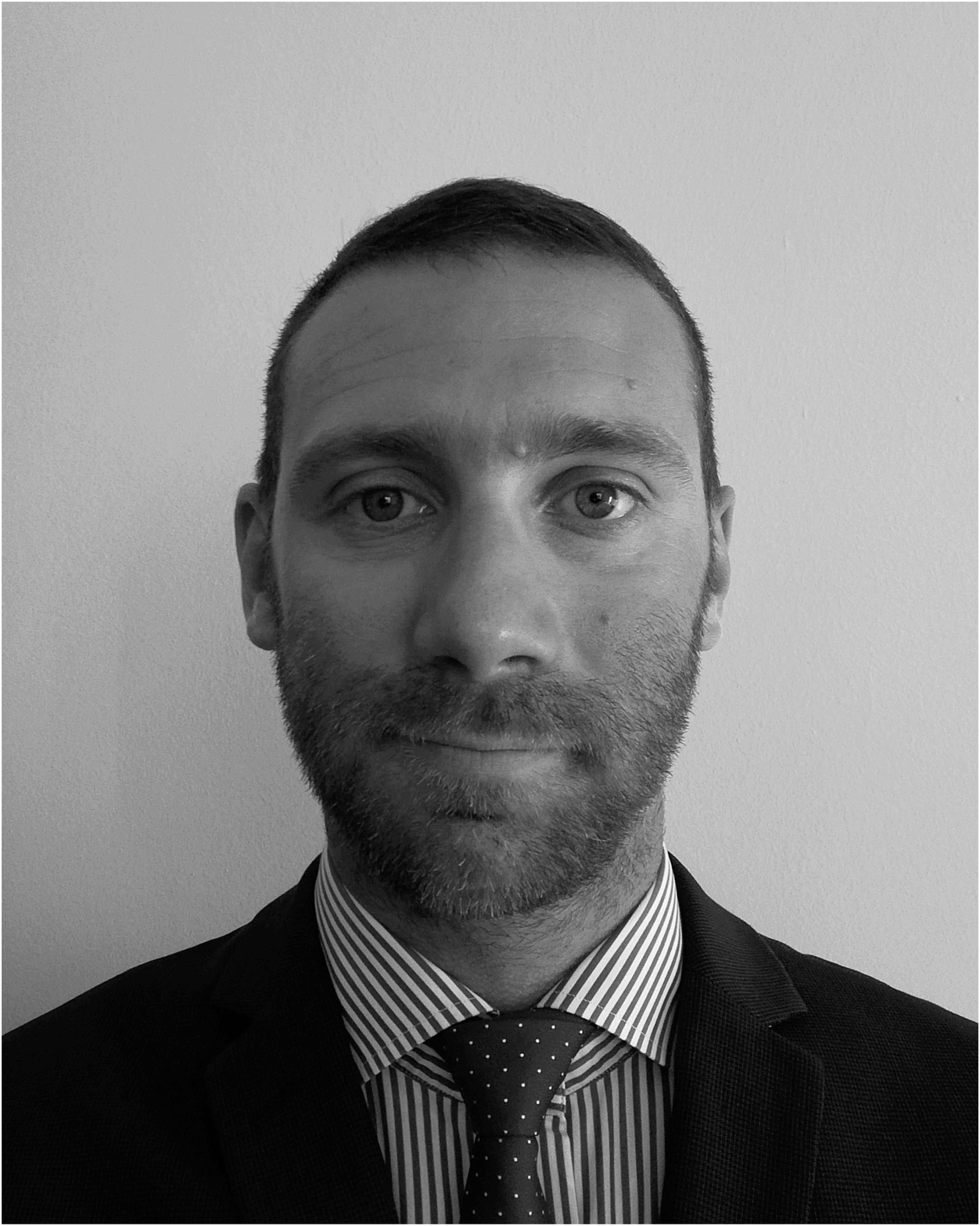}}]{Fabio D'Agostino}(S’13 – M’16) received the master’s degree, and the Ph.D. in Electrical Engineering from the University of Genova, in 2013 and 2017. He is currently Assistant Professor at the Department of Electrical, Electronic, Telecommunication Engineering and Naval Architecture of the University of Genova. His current research activity includes protection and control of distribution networks, microgrids and shipboard power systems. From 2021 he is one of the representatives of the IEEE Marine Systems Coordinating Committee (MSCC), and he is the Secretary of the CIGRE Working Group C1.45. In 2015 he has been a Visiting Scholar at the Electrical Engineering and Computer Science department of the Washington State University, where he worked in the distribution system reliability area.
\end{IEEEbiography}

\begin{IEEEbiography}[{\includegraphics[width=1in,height=1.25in,clip,keepaspectratio]{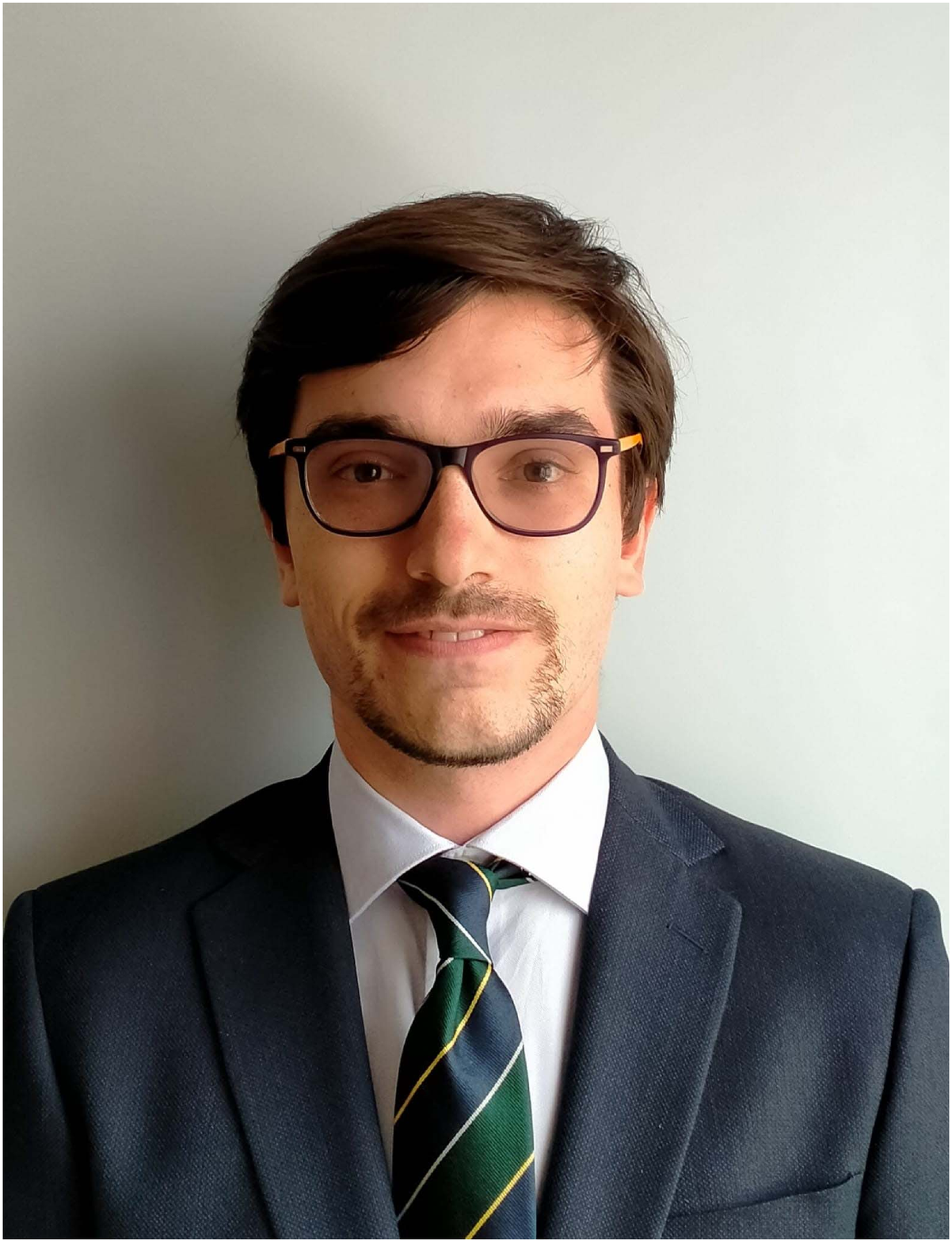}}]{Bruno Gabriele}(S'19) was born in Genoa, in Italy, in 1995. He received the B.Sc. and M.Sc. degrees in electrical engineering from the University of Genoa, Italy, in 2017 and 2019, respectively. He is currently a Ph.D. student in Sciences and Technologies for Electrical Engineering and Complex Systems for Mobility at the University of Genoa. His research interests include the evolution of the power systems and electricity markets and the integration of renewable energy sources.
\end{IEEEbiography}

\begin{IEEEbiography}[{\includegraphics[width=1in,height=1.25in,clip,keepaspectratio]{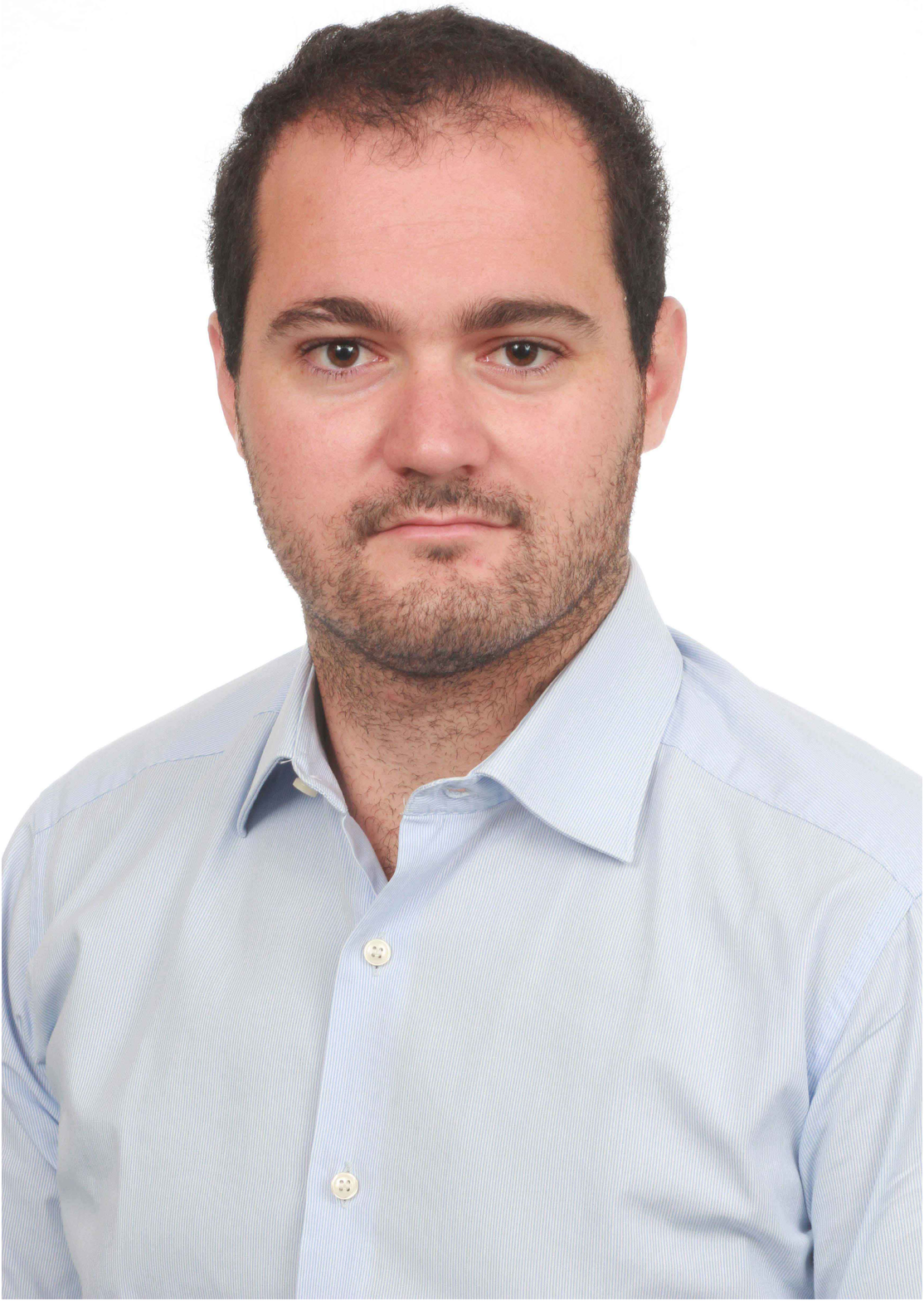}}]{Giacomo-Piero Schiapparelli}(S'18) was born in Genoa in 1994. He got B.Sc and M.Sc (Hons.) degree in electrical engineering from the University of Genoa. In 2021 he earned his International PhD in Science and Technology for Electrical Engineering, Marine Engineering, Complex Systems for Mobility, curriculum: Electrical Engineering at the University of Genoa. He published more than 15 peer-reviewed research papers in high-level international conferences and journals during his academic career. He spent one year doing research abroad at the École Polytechnique Fédérale de Lausanne (EPFL). He is currently responsible for system engineering and power electronics at High-Performance Engineering Srl (HPE). He is researching and developing products in the electrification of automotive, motorsport, off-highway sectors. Moreover, Dr Schiapparelli is also involved in the TRANSFORM project (H2020-ECSEL-2020-1-IAtwo-stage) as task leader (WP6 T6.3) and steering committee member.
\end{IEEEbiography}

\begin{IEEEbiography}[{\includegraphics[width=1in,height=1.25in,clip,keepaspectratio]{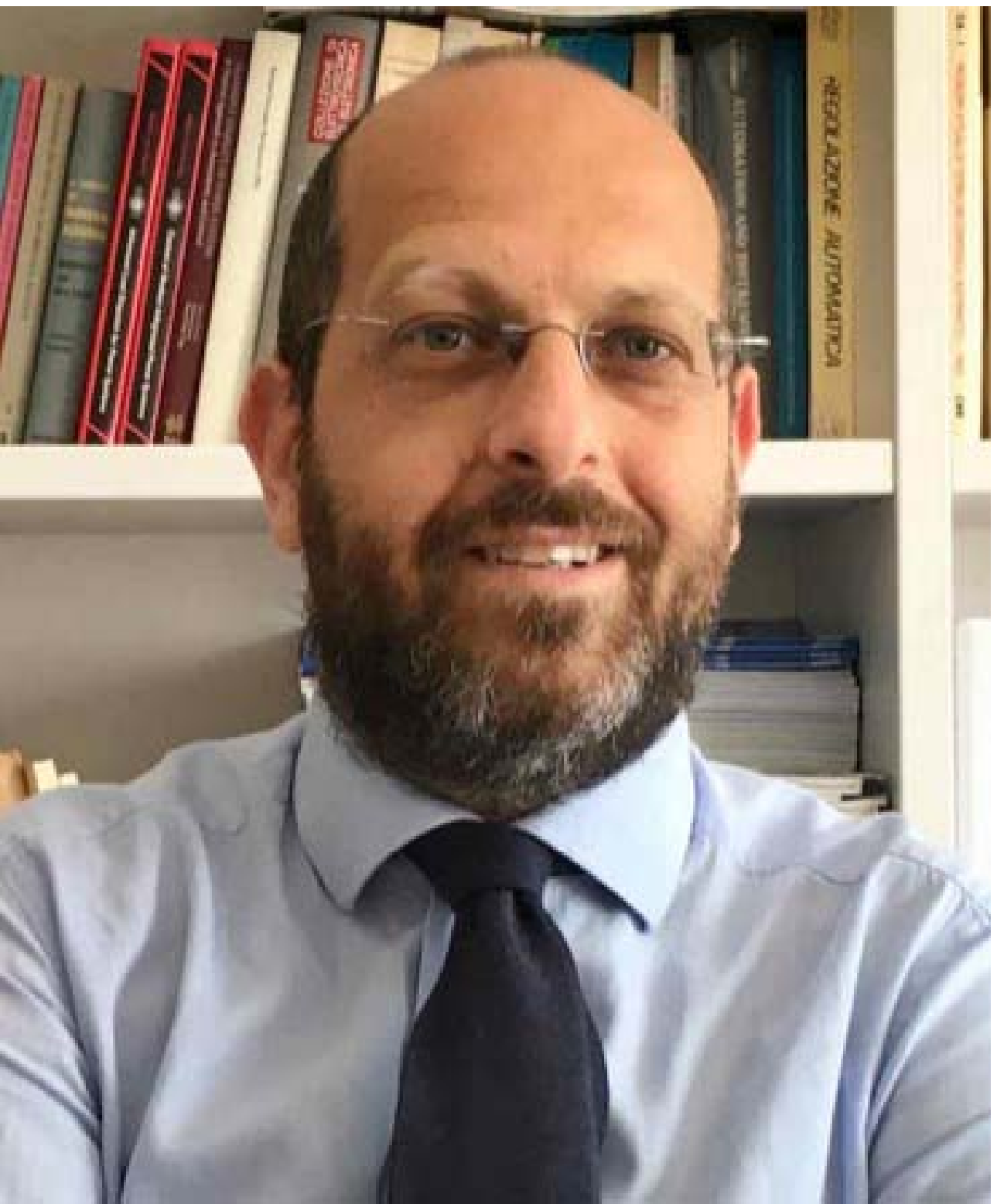}}]{Federico Silvestro} (S'01 - M'02 - SM'16) was born in Genoa, Italy, in 1973. He received the electrical engineering degree and the Ph.D. degree in electric power systems from the University of Genoa in 1998 and 2002, respectively.

He is currently Full Professor at University of Genova - DITEN (Department of Electrical, Electronics and Telecommunication Engineering and Naval Architecture). Prof. Silvestro has been scientific responsible for different research projects. He has authored over 200 scientific papers. His current research interests include power system optimization and microgrids, energy saving, dynamic security assessment, marine application and knowledge-based systems applied to power systems.

Prof. Silvestro is currently secretary of the CIGR\'E Working Group JWG C1/C6.37/CIRED on Optimal Transmission and Distribution Investment Decisions Under Increasing Energy Scenario Uncertainty.
\end{IEEEbiography}
\vspace{200pt}

\end{document}